\documentclass[aps, prd, superscriptaddress, nofootinbib, preprintnumbers]{revtex4-2}
\usepackage[colorlinks=true]{hyperref}
\usepackage{xcolor}
\hypersetup{colorlinks=true, citecolor=blue, urlcolor=blue, linkcolor=blue}
\usepackage{amsmath,amssymb}
\usepackage[final]{graphicx}
\usepackage{subcaption}
\usepackage[belowskip=0.5pt,aboveskip=0.5pt]{caption}
\usepackage{float}
\usepackage{adjustbox}
\usepackage{cancel,soul,ulem}

\begin{document}

\title{The effect of charm quark on chiral phase transition in $N_f=2+1+1$ holographic QCD}
\author{Hiwa A. Ahmed}
\thanks{{\tt hiwa.a.ahmed@mails.ucas.ac.cn}}
\affiliation{School of Nuclear Science and Technology,
University of Chinese Academy of Sciences, Beijing 100049, China}
\affiliation{Department of Physics, College of Science, Charmo University, 46023, Chamchamal, Sulaymaniyah, Iraq}
\author{Mamiya Kawaguchi}\thanks{{\tt mamiya@ucas.ac.cn}} 
\affiliation{School of Nuclear Science and Technology,
University of Chinese Academy of Sciences, Beijing 100049, China}
\author{Mei Huang}\thanks{{\tt huangmei@ucas.ac.cn}}
\affiliation{School of Nuclear Science and Technology, University of Chinese Academy of Sciences, Beijing 100049, China}

\date{\today}

\begin{abstract}
We investigate the effect of charm quark on the chiral phase transition of light quarks at finite temperature based on the four-flavor soft-wall holographic QCD model. In the massless limit, we find that the thermal chiral phase transition is of the second order in the four-quark flavor system. In the case with the massive charm quark and the massless light and strange quarks, the order of the phase transition changes to the first order. This is due to the quark flavor symmetry breaking which is associated with the violation of the $U(1)$ axial symmetry.
Once the light and strange quarks get massive, the explicit chiral symmetry breaking becomes eminent,  then the crossover phase transition is realized at the physical quark masses.
We also map the order of the phase transition on a phase diagram in the quark mass plane where the light- and strange-quark masses are degenerate but differ from the value of the charm quark mass.
This phase diagram is an extension of the conventional Columbia plot to the four-quark flavor system.
The critical exponents related to the chiral phase transition are also addressed. 

\end{abstract}


\maketitle
%%%%%%%%%%%%%%%%%%%%%%%%%%%%%%%%%%%%%%%%%%%%%%%%%%%%%%%%%%%%%%%%%%%%%%%%
\section{Introduction}
\label{sec-Intro}
Investigating the thermal QCD phase transition is an important subject for understanding the hot environment relevant to the initial stage of heavy-ion collisions and the early universe. 
To clarify its phase structure,
the order of the chiral phase transition, especially depending on the number of quark flavors ($N_f$) and the current quark masses ($m_f$), has been extensively investigated by the lattice QCD framework~\cite{Brown:1990ev,Aoki:2012yj,Bazavov:2017xul,HotQCD:2019xnw,Ding:2019fzc,Kuramashi:2020meg,Dini:2021hug}, the functional renormalization group (FRG) method~\cite{Braun:2006jd,Braun:2009ns,Fukushima:2010ji,Grahl:2013pba,Fejos:2014qga,Resch:2017vjs,Fejos:2022mso,Braun:2023qak}, Dyson-Schwinger equations~\cite{Gao:2021vsf,Bernhardt:2023hpr} and chiral effective model approaches~\cite{Pisarski:1983ms,Lenaghan:2000kr,Fukushima:2008wg}.

In a pioneer work, the existence of universality classes has been investigated by using the renormalization group analysis in the linear sigma model approach at the one-loop calculation with the $\epsilon$-expansion. 
They have concluded that the chiral phase transition is of the first order for $N_{f} \ge 3$ with massless quarks~\cite{Pisarski:1983ms}.
Motivated by this first research, the order of the chiral phase transition in the three quark flavor system ($N_{f} = 3$) has been mapped onto the plane of the light- and strange-quark masses. This phase diagram is referred to as the Columbia plot~\cite{Brown:1990ev}. 
Using effective model approaches with the three quark flavors, it has been suggested that a domain of the first order phase transition emerges around the small quark mass regions~\cite{Pisarski:1983ms,Lenaghan:2000kr,Fukushima:2008wg}.
These model analyses imply that 
the emergence of the chiral first order phase transition is closely linked to the $U(1)$ axial anomaly. 
However, the definite evidence of the first order domain has not been observed in lattice QCD simulations.
The specifics of the Columbia plot remains unclear and the details of the chiral phase transition depending on the quark flavors are still not well understood.

Shifting our focus to the multi-flavor system ($N_f\geq 4$) could offer a fresh perspective on the thermal QCD phase transition.
Recently,
the lattice QCD study has endeavored to extend 
the Columbia plot to the multi-flavor system
incorporating more than four quark flavors~\cite{Cuteri:2021ikv}.
According to this lattice simulation,
the chiral phase transition is estimated to be of the second order with $N_{f}\le 6$ after extrapolation to the massless limit.
This lattice estimation actually differs from the first study based on searching the universality class in the linear sigma model approach~\cite{Pisarski:1983ms}.
Investigating the chiral phase transition in the multi-flavor system has thus more clearly highlighted 
the distinction between the lattice QCD observation and the effective model results.

Although the extension has been made by the recent lattice work~\cite{Cuteri:2021ikv}, our understanding of the phase structure in the multi-flavor system 
is still limited even compared to the study of the conventional Columbia plot described with the light and strange flavors.
In our study, 
we address the thermal chiral phase transition in such a multi-flavor system, specifically focusing on the four-quark flavor system, by using the holographic QCD approach, and investigate details such as critical exponents related to the chiral phase transition with $N_f = 4$.

The anti-de Sitter/conformal field theory (AdS/CFT) correspondence and the conjecture of the gravity/gauge duality is a powerful tool for investigating the non-perturbative QCD property~\cite{Maldacena:1997re,Witten:1998qj}. 
There are two main approaches in holographic QCD: the top-down approach and the bottom-up approach.
The top-down approach begins with a string theory in high dimensions and attempts to create a low-dimensional theory with characteristics similar to QCD. On the other hand, the bottom-up approach starts with established QCD phenomenology and tries to understand the features that its high-dimensional gravitational dual exhibits. 
As a development of the bottom-up approach,
%it has been pointed out that 
the dilaton configuration plays an important role in providing the realistic QCD phenomena \cite{Karch:2006pv,Chelabi:2015cwn,Chelabi:2015gpc}:
taking into account the infrared (IR) behavior of the dilaton, the soft-wall AdS/QCD model can successfully reproduce the linear behavior of meson spectra~\cite{Karch:2006pv}. Furthermore, the spontaneous chiral symmetry breaking in the chiral limit is well described by the ultraviolet (UV) behavior of the dilaton~\cite{Chelabi:2015cwn,Chelabi:2015gpc,Li:2016smq}.
In this study, we use the bottom-up holographic QCD, which has been employed to explain the QCD phenomena observed in experimental physics.

With the developed soft-wall model,  
the phase structure has been studied at finite temperature and/or finite chemical potentials~\cite{Colangelo:2011sr,Cui:2013zha,Bartz:2016ufc,Chelabi:2015cwn,Chelabi:2015gpc,Fang:2015ytf,Fang:2016nfj,Li:2016smq,Bartz:2017jku}.   
The Columbia plot has also been described by the previous study in the holographic approach. 
In our study, we extend the previous works of the soft-wall model to the four-quark flavor system at finite temperature and propose the extended Columbia plot.
To characterize the phase structure with $N_f = 4$, 
we also investigate the critical exponents associated with the order parameter of the chiral phase transition.

The paper is organized as follows. Firstly, we will provide a brief introduction about the soft wall AdS/QCD models with $N_f=4$ and present the computation of the order parameter (quark condensate) in section \ref{sectionII}. Then, in section \ref{sectionIII}, we will study the order of the chiral phase transition for the two-, three- and four-quark flavor systems.
 Additionally, we will map the order of the chiral phase transition in the three quark flavor system onto the plane of the light- and strange-quark masses, and we will investigate the charm quark mass contribution. Towards the end of section \ref{sectionIII}, we will reach the main goal of our work, which is the extension of the conventional Colombia plot in the three-quark flavor system to the four-quark flavor system where the light- and strange-quark masses are degenerate but differ from the value of the charm quark mass.
 In this section, we will also address the critical exponents related to the chiral phase transition. 
 Finally, we will summarize and discuss the results in section \ref{sectionIIII}.

%%%%%%%%%%%%%%%%%%%%%%%%%%%%%%%%%%%%%%%%%%%%%%%%%%%%%%%%%%%%%%%%%%%%%%%%
%%%%%%%%%%%%%%%%%%%%%%%%%%%%%%%%%%%%%%%%%%%%%%%%%%%%%%%%%%%%%%%%%%%%%%%%
%%%%%%%%%%%%%%%%%%%%%%%%%%%%%%%%%%%%%%%%%%%%%%%%%%%%%%%%%%%%%%%%%%%%%%%%

\section{$N_f$-quark flavor system based on holographic QCD approach}
\label{sectionII}
Our aim of this paper is to investigate 
the quark-flavor dependence on 
the chiral phase diagram based on the holographic QCD approach. 
In this section, we begin by giving a brief review of the soft-wall model which is applicable to 
the chiral phase transition at finite temperature.
This model is defined in the AdS$_5$ space-time.
The temperature is introduced through the geometries with black holes
and it is described by the background metric where there is no dynamical contribution. In our work, we take the solution of the simple AdS-Schwarzschild (AdS-SW) black hole: 
\begin{equation}
ds^{2}= \frac{L^2}{z^2} \left( - f(z) dt^{2} + \frac{1}{f(z)} dz^{2} + dx_{i} dx^{i}\right),
\end{equation}
where
$z$ represents the fifth coordinate and $L$ the radius of the AdS curvature ($L$ is set to $1$ below for simplicity).
The function $f(z)$ is given by
\begin{equation}
f(z) = 1- \frac{z^{4}}{z_{h}^{4}},
\end{equation}
where $z_{h}$ denotes the black hole horizon defined at $f(z_{h})=0$. 
Using this function $f(z)$, we have the Hawking temperature formula at the black hole horizon,

\begin{equation}
    T=\frac{1
    }{4 \pi}
    \left|\frac{d f(z)}{dz}
    \right|_{z \to z_{h} } = \frac{1}{\pi z_{h}}.
\end{equation}
This $T$ is identified as the temperature of the hot QCD system within the holographic model approach.

%%%%%%%%%%%%%%%%%%%%%%%%%%%%%%%%%%%%%%%%%%%%%%%%%%%%%%%%%%%
\subsection{Soft-wall model with multi-quark flavors}

Following the original work of the bottom-up holographic QCD model \cite{Karch:2006pv}, 
we employ a five-dimensional gauge model respecting the $U(N_f)_L\times U(N_f)_R$ symmetry with $N_f$ being the number of the quark flavors.
The five-dimensional model is described by the bulk left- ($L_M$) and right- ($R_M$) gauge fields. 
To incorporate the quark condensation operator in the bottom-up approach, 
we introduce a bulk scalar field $X$ which
is a $N_f\times N_f$ matrix and transforms as bifundamental representation under the $U(N_f)_L\times U(N_f)_R$ symmetry.
Furthermore, considering  the breaking of the conformal invariance, 
we also include the dilaton background field $\Phi(z)$ which acts as a smooth cutoff. 
The five-dimensional soft-wall action takes the form:

\begin{equation}
\begin{aligned}
S_{M} &=-\int d^{5} x \sqrt{-g} e^{-\Phi}\left\{ \operatorname{Tr}\left[\left(D^{M} X\right)^{\dagger}\left(D_{M} X\right) +\frac{1}{4 g_{5}^{2}}\left(L^{M N} L_{M N}+R^{M N} R_{M N}\right)\right]
+V(X)
\right\},
\end{aligned}
\label{action}
\end{equation}
where
$g$ denotes the determinant of metric $g_{MN}$: $\sqrt{-g} = (1/z)^5$;
the covariant derivative of $X$ is given by
$D^{M} X=\partial_{M} X - i L_{M} X + i X R_{M}$;
$L_{MN}$ and $R_{MN}$ are the field strength 
defined as $L_{MN}=\partial_M L_N -\partial_N L_M -i[L_N,L_M]$
and $R_{MN}=\partial_M R_N -\partial_N R_M -i[R_N,R_M]$ in terms of the left and right hand gauge field respectively;
$g_{5}$ denotes the gauge coupling constant
determined by matching with the UV asymptotic behavior of the vector current correlator $g_{5}^{2} = 12 \pi^{2} /N_{c}$ \cite{Erlich:2005qh}.
$V(X)$ represents the scalar potential 
which takes the polynomial form of $X$:
\begin{eqnarray}
V(X) = V_0(X) + V_{\rm det}.
\end{eqnarray}
$V_0$ is constructed by only the quark-flavor symmetric interactions, which is actually invariant under the $U(N_f)_L\times U(N_f)_R$ symmetry,
\begin{eqnarray}
V_0(X)= M_{5}^{2}{\rm Tr}\left[ X^{\dagger} X\right] 
+\lambda {\rm Tr}\left[ |X|^{4}\right]
\end{eqnarray}
where the mass-parameter $M_{5}^{2}$ 
is given by $M_{5}^{2}=(\Delta - p)(\Delta + p -4) =-3$ with $\Delta=3$ and $p=0$ through the holographic dictionary, 
and $\lambda$ is a positive-dimensionless quartic coupling.
In the absence of the $X^4$ term, 
the quark condensate is linked with the current quark mass.
At the chiral limit, the condensate accidentally vanishes. To avoid this issue, the $X^4$ term is included in the potential $V_0(X)$ \cite{Gherghetta:2009ac}.

In addition, the potential part $V(X)$ also has the determinant term as in Ref. \cite{Eshraim:2014eka} to provide the quark-flavor mixing term,
\begin{eqnarray}
V_{\rm det}= \gamma {\rm Re}\,[{\rm det}(X)],
\end{eqnarray}
with $\gamma$ being a real parameter.
Note that this determinant term is anomalous under the $U(1)_A$ symmetry, but it is invariant under the $SU(N_f)_L\times SU(N_f)_R$.

In this study, we mainly focus on the four flavor system. Hence, the following analysis is based on the soft-wall model with $N_f =4$.

To determine the classical solution of $X$, 
we consider the static homogeneous expectation value of the scalar field:

\begin{equation}
X_0=
{\rm diag} \left( \frac{\chi_u(z)}{\sqrt{2}},
\frac{\chi_d(z)}{\sqrt{2}},
\frac{\chi_s(z)}{\sqrt{2}},
\frac{\chi_c(z)}{\sqrt{2}}\right)
\label{chivev}
\end{equation}
where the factor $\frac{1}{\sqrt{2}}$ is taken to get a canonical form of the kinetic terms of 
$\chi_{f}$ ($f=u,d,s,c$). 
In this study, we impose the isospin symmetry, $m_l=m_u= m_d$, so that the light sector of the scalar field $X_0$ takes $\chi_l = \chi_u = \chi_d$ through the matching condition with QCD.  
Since we are concerned with the chiral phase transition,
we set the bulk gauge fields $L_M$ and $R_M$ to zero.
With this setup, 
the action in Eq.~(\ref{action}) is evaluated as
\begin{equation}
\begin{aligned}
S\left[\chi_l, \chi_s, \chi_c\right]= & -\int d^5 x \sqrt{-g} e^{-\Phi} \left\{g^{z z}\left(\chi_l^{\prime 2}+\frac{1}{2} \chi_s^{\prime 2}  +\frac{1}{2} \chi_c^{\prime 2}\right) \right. \\
&  \left. + \left[-3\left(\chi_l^2+\frac{1}{2} \chi_s^2 +\frac{1}{2} \chi_c^2\right)+v_4\left(2 \chi_l^4+\chi_s^4 +\chi_c^4\right) +3 v_{\rm det} \chi_l^2 \chi_s \chi_c\right]\right\}
\end{aligned}
\label{actionchi}
\end{equation}
with 
$v_{\rm det}= \gamma/12$ and $v_{4}=\lambda/4$.

From this action, the equations of motion for $\chi_f$ ($f=l,s,c$) read
\begin{equation}
\begin{gathered}
\chi_l^{\prime \prime}+\left(-\frac{3}{z}-\Phi^{\prime}+\frac{f^{\prime}}{f}\right) \chi_l^{\prime}+\frac{1}{z^2 f}\left(3 \chi_l-3 v_{\rm det} \chi_l \chi_s \chi_c- 4 v_4 \chi_l^3\right)=0, \\
\chi_s^{\prime \prime}+\left(-\frac{3}{z}-\Phi^{\prime}+\frac{f^{\prime}}{f}\right) \chi_s^{\prime}+\frac{1}{z^2 f}\left(3 \chi_s-3 v_{\rm det} \chi_l^2 \chi_c -4 v_4 \chi_s^3\right)=0, \\
\chi_c^{\prime \prime}+\left(-\frac{3}{z}-\Phi^{\prime}+\frac{f^{\prime}}{f}\right) \chi_c^{\prime}+\frac{e^{1}}{z^2 f}\left(3 \chi_c-3 v_{\rm det} \chi_l^2 \chi_s-4 v_4 \chi_c^3\right)=0, \\
\end{gathered}
\label{chieom}
\end{equation}
where the prime denotes the derivative with respect to $z$ component.
These equations are a set of coupled equations, which is   
driven by the determinant term associated  with  
the quark-flavor violation ($m_l\neq m_s\neq m_c$).
%the $U(1)_A$ anomalous term 
Indeed, in the four quark flavor symmetric limit, $m_l=m_s=m_c$, 
the three coupled equations reduce to a single equation.

In the soft wall holographic QCD model, a non-dynamical background field called dilaton field $\Phi$, which is only a function of $z$ added to break the conformality of the AdS space. This background field plays an important role in the phenomenological observation and the chiral dynamics. Since the dilaton modifies the equations of motion, it affects mesons
spectra, form factors, and any other property derived from the action. 
A positive dilaton field at the infrared (IR) region $\phi(z \to \infty) \approx +z^2$ is necessary to get the linear confinement~\cite{Karch:2006pv}. 
In contrast to IR region, the dilaton field has to be a negative quadratic $\phi(z \to 0) \approx -z^2$
in the ultraviolet (UV) limit to describe the spontaneous symmetry breaking in the chiral limit, as pointed out in Ref. \cite{Chelabi:2015cwn,Chelabi:2015gpc}. The specific profile of the dilaton field, which respect both the IR and UV behavior, is proposed as \cite{Chelabi:2015cwn}

\begin{equation}
    \Phi(z)=- \mu_{1}^{2}z^2  + (\mu_{1}^{2} + \mu_{0}^{2}) z^{2} \text{tanh}(\mu_{2}^{2} z^{2}),
\end{equation}
where the parameters $\mu_{0}=0.43$ GeV is fixed by the radial excitation of the $\rho$ meson masses, and  $\mu_{1}=0.83$ GeV and $\mu_{2}=0.176$ GeV are chosen to get 
the critical temperature of the second order
phase transition around $150$~MeV for $N_f=2$ with massless quarks~\cite{Chelabi:2015cwn,Chelabi:2015gpc}.

%%%%%%%%%%%%%%%%%%%%%%%%%%%%%%%%%%%%%%%%%%%%%%%%%%%%%%%%%%%%%%%%%%%%
%%%%%%%%%%%%%%%%%%%%%%%%%%%%%%%%%%%%%%%%%%%%%%%%%%%%%%%%%%%%%%%%%%%%

\subsection{Matching condition with QCD and numerical demonstration}

Using the holographic correspondence,
the quark condensate in QCD is extracted 
from the UV asymptotic behavior of the scalar fields $\chi_f$  ($f=l,s,c$) in the five dimensional model~\cite{Erlich:2005qh,Karch:2006pv},
which serves as the order parameter of the spontaneous chiral symmetry breaking. 
Indeed, $X/z$ is dual to the $\bar qq$ operator within the holographic QCD approach.
Through the matching condition with QCD at the UV ($z\to 0$), $X/z$ is identified as the current quark mass $m_f$ corresponding to the source of the quark condensate.

The UV solutions of $\chi_f$ can be obtained by using the Frobenius method near the boundary $z=0$~\cite{Li:2020hau}. 
At small $z$, the general expression of $\chi_f$ are given by
\begin{equation}
\begin{gathered}
    \chi_{l} = a_{l} z - \left( \mu_{1}^{2} - a_{l}^{2} v_{4} - \frac{3}{2} a_{s} a_{c} v_{det}  \right) a_{l} z^{3} \log(z) + b_{l} z^{3} + ..... ,\\
    \chi_{s} = a_{s} z - \left( \mu_{1}^{2} a_{s} - a_{s}^{3} v_{4} - \frac{3}{2} a_{l}^{2} a_{c} v_{det}  \right)  z^{3} \log(z) + b_{s} z^{3} + ..... ,\\
    \chi_{c} = a_{c} z - \left( \mu_{1}^{2} a_{c} - a_{c}^{3} v_{4} - \frac{3}{2} a_{l}^{2} a_{s} v_{det}  \right)  z^{3} \log(z) + b_{c} z^{3} + ..... ,\\
\end{gathered}
\label{UV}
\end{equation}
where 
$a_f$ and $b_f$ are the integral constants of the three coupled differential equations.
Owing to the holographic dual of QCD at the UV boundary, 
the coefficients of the $z$ terms in
Eq.~(\ref{UV}) correspond to the current quark masses:
\begin{equation} 
\begin{aligned}
 a_{l}= m_{l} \zeta, \quad a_{s}= m_{s} \zeta, \quad a_{c}= m_{c} \zeta. 
\end{aligned}
\end{equation}
with $\zeta=\frac{\sqrt{N_c}}{2\pi}$~\cite{Cherman:2008eh}. 
By considering the linear response with respect to the current quark masses, the coefficients of the $z^3$ terms in Eq.~(\ref{UV}) are identified as
the quark condensates, $\sigma_l\equiv \langle \bar u u\rangle=\langle \bar d d\rangle$, $\sigma_s\equiv \langle \bar ss\rangle$ and $\sigma_c\equiv \langle \bar cc\rangle$:
\begin{equation}
\begin{aligned}
 b_{l}= \frac{\sigma_{l}}{\zeta}, \quad b_{s}= \frac{\sigma_{s}}{\zeta}, \quad b_{c}= \frac{\sigma_{c}}{\zeta}
\end{aligned}.
\end{equation} 
In the numerical calculation, the integral constants $a_f$ are treated as input parameters with the values of quark masses. On the other hand, 
$b_f$ linked with quark condensates
are determined by solving the equations of motion in Eq.~(\ref{chieom}) with appropriate boundary conditions.

To numerically solve the equations of motion, we first impose   
the boundary condition near the UV, 

\begin{equation}
    \lim_{\epsilon \to 0} \frac{\chi_{l}(\epsilon)}{\epsilon}= m_{l} \zeta, \quad \lim_{\epsilon \to 0} \frac{\chi_{s}(\epsilon)}{\epsilon}= m_{s} \zeta, \quad \lim_{\epsilon \to 0} \frac{\chi_{c}(\epsilon)}{\epsilon}= m_{c} \zeta,
\label{UVbound}
\end{equation}
where $\epsilon$ is a small number. In addition, another boundary condition arises from 
the black hole horizon defined at $f(z_{h})=0$.
In fact, 
the coupled equations of motion 
in Eq.~(\ref{chieom}) have a singularity at the horizon $z=z_h$, resulting in 
the unexpected divergence of $\chi_f$. 
To avoid the singularity at the horizon $z=z_h$, the following boundary condition should be applied to numerically solve the equations of motion, 
\begin{equation}
\begin{gathered}
f^{\prime} \chi_l^{\prime}+\frac{1}{z^2}\left(3 \chi_l-3 v_{\rm det} \chi_l \chi_s \chi_c- 4 v_4 \chi_l^3\right)|_{z=z_h-\epsilon}=0, \\
f^{\prime} \chi_s^{\prime}+\frac{1}{z^2}\left(3 \chi_s-3 v_{\rm det} \chi_l^2 \chi_c -4 v_4 \chi_s^3\right)|_{z=z_h-\epsilon}=0, \\
f^{\prime} \chi_c^{\prime}+\frac{1}{z^2}\left(3 \chi_c-3 v_{\rm det} \chi_l^2 \chi_s-4 v_4 \chi_c^3\right)|_{z=z_h-\epsilon}=0.\\
\end{gathered}
\label{horizon}
\end{equation}
With these boundary conditions, we perform numerical calculations in several cases to provide practical examples, below.

\begin{figure}
\begin{subfigure}{0.49\textwidth}
  \centering
  \includegraphics[width=1\linewidth]{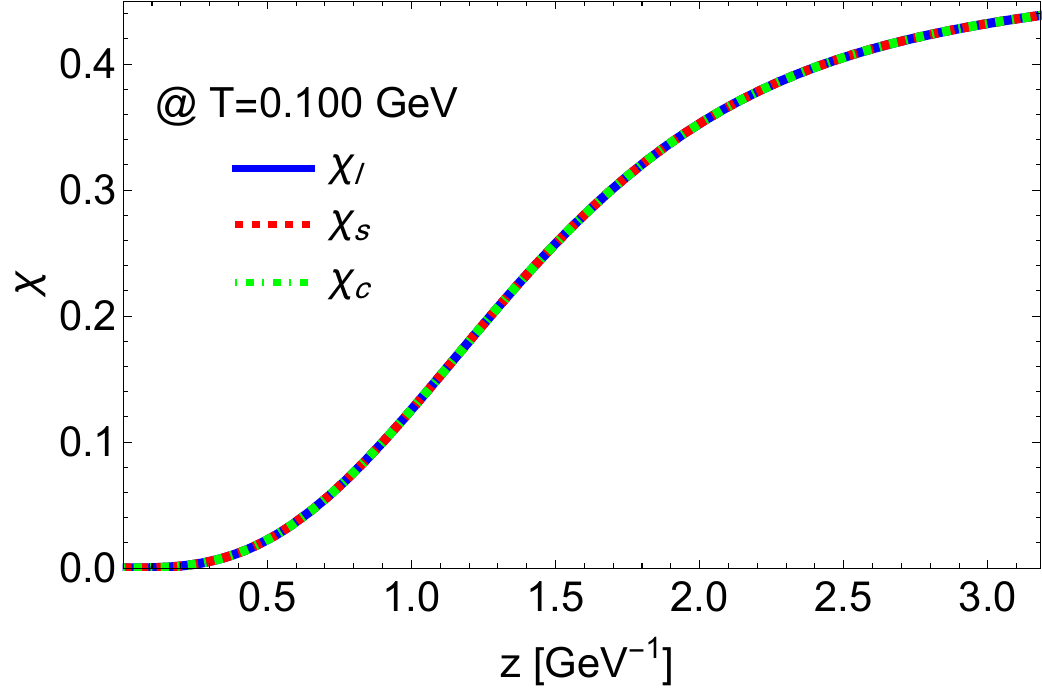} 
   \caption{ }
\label{chisolutionza}
\end{subfigure} 
\begin{subfigure}{0.49\textwidth}
  \centering
  \includegraphics[width=1\linewidth]{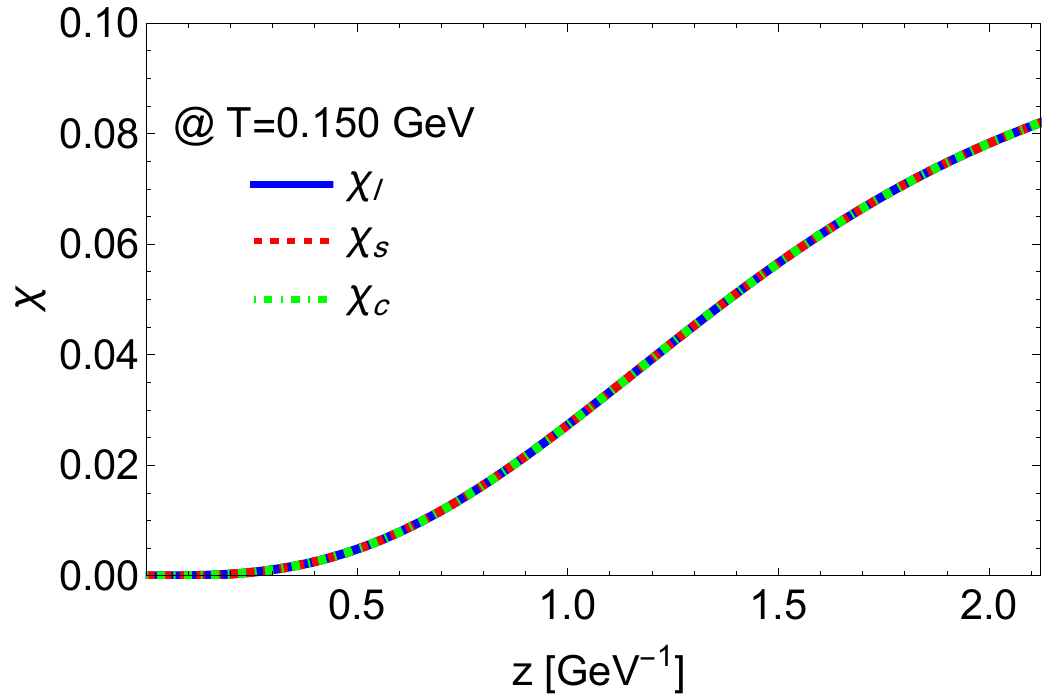}
 \caption{ }
\label{chisolutionzb}
\end{subfigure}%
\caption{Solutions of $\chi_l$, $\chi_s$, and $\chi_c$ as functions of holographic coordinate $z$ at massless limit $m_{l}=m_{s}=m_{c}=0$ for $T=0.100$ GeV (a) and $T=0.150$ GeV (b). }
\label{chisolutionz}
\end{figure}

The coupled equations of motion in Eq. \eqref{chieom} can be solved by utilizing the shooting method with the boundary conditions specified in Eqs.~\eqref{UVbound} and \eqref{horizon} (the outlined of the shooting method is given in Refs. \cite{Chelabi:2015cwn,Bartz:2017jku}). By shooting from the near horizon limit $z=z_h-\epsilon$ to the near UV boundary $z=\epsilon$, 
we obtain the numerical solutions of $\chi_f$ 
for the case of the four-flavor symmetric massless limit, 
$m_{l}=m_{s}=m_{c}=0$, in Fig.~\ref{chisolutionz} where we take $T=0.100$~GeV and $T=0.150$~GeV corresponding to $z_h \simeq 3.18\,{\rm GeV}^{-1}$ and $z_h \simeq 2.12\,{\rm GeV}^{-1}$ respectively. 
This figure shows that the numerical solutions of $\chi_f$ surely satisfy the boundary conditions and there is no singularity.
Note that the solutions of the scalar fields result in $\chi_l=\chi_s=\chi_c$, owing to the four-flavor symmetry, which implies that the quark condensates are equal for each other, $\sigma_{l}=\sigma_{s}=\sigma_{c}$. 
The value of the parameter $v_{4}$ is fixed at $v_{4}=8$ by following 
the parameter set in the previous analyses based on the soft-wall model with $N_f=2$ and $N_f=3$
~\cite{Chelabi:2015cwn,Chelabi:2015gpc}, which is also similar to the values of the quartic term couplings presented in Ref.~\cite{Ahmed:2023zkk}. For $N_f=3$, the $v_{\rm det}$ parameters is defined by the coupling of the determinant term as $v_{det}= \frac{\gamma}{6\sqrt{2}}$ and assigned to be $-3$ \cite{Chelabi:2015cwn,Chelabi:2015gpc}. 
As for $N_f=4$,
the parameter $v_{\rm det}$ is given by $v_{\rm det}= \frac{\gamma}{12}$ . Since one can go from 
the four-quark flavor system to the three-quark flavor system by integrating out the contribution of the charm quark with the heavy quark limit, $m_{c}\to \infty$, 
the parameter $v_{\rm det}$ is redefined 
as $v_{det}=\gamma/(12 ~ \chi_c(z_{h}) |_{m_c\to \infty})$.
To match the value in the three-flavor system, $v_{\rm det}$ in the four-flavor system is assigned to be $-7.07$.

The quark condensate $\sigma_f$ is determined by fitting the UV behavior of $\chi_f$ in Eq.~\eqref{UV}. 
In the case of the massless quarks, the solution of $\sigma_f$ is identical for all flavors, resulting in equal condensates, $\sigma_{l}=\sigma_{s}=\sigma_{c}$. The values of the condensates for $T=0.100$~GeV and $T=0.150$~GeV are evaluated as $\sigma_{l}=\sigma_{s}=\sigma_{c}=(0.3794 \text{ GeV})^{3}$ and $\sigma_{l}=\sigma_{s}=\sigma_{c}=(0.2276 \text{ GeV})^{3}$, respectively.

\begin{figure}
  \centering
  \includegraphics[width=0.5\linewidth]{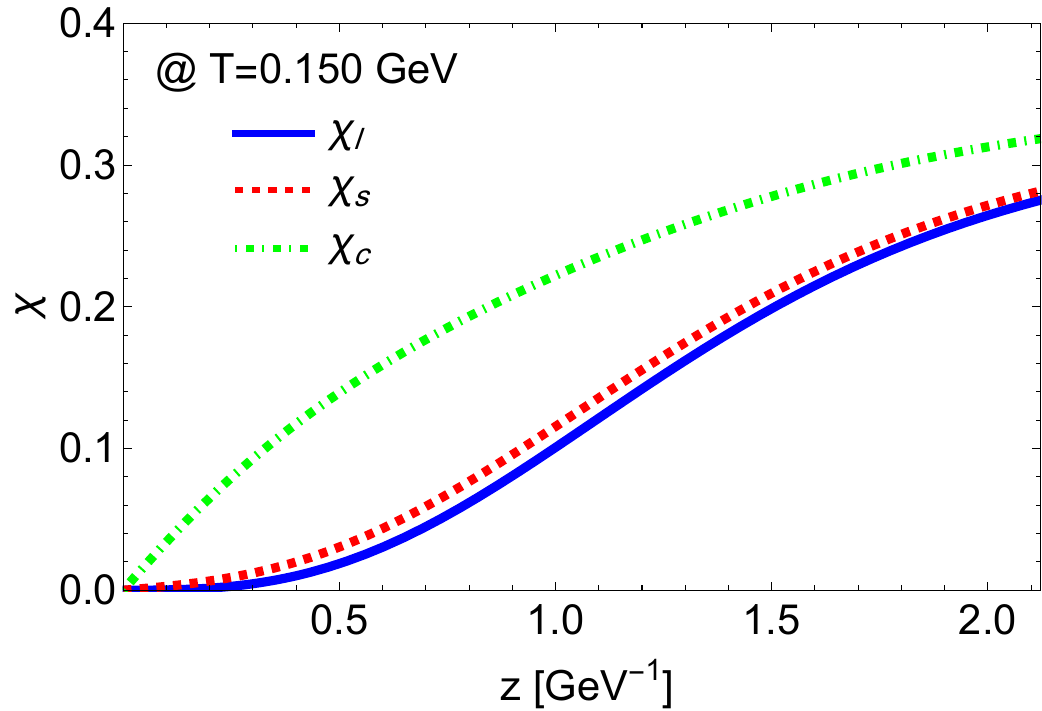}   

\caption{Solutions of $\chi_l$, $\chi_s$, and $\chi_c$ as functions of $z$ at the physical quark masses $m_{l}=3.45 \text{ MeV}$, $m_{s}=93.4 \text{ MeV}$, and $m_{c}=1.27 \text{ GeV}$ and temperature $T=0.150$ GeV.}
\label{chisolutionzphys}
\end{figure}

Similarly, the equation of motions for $\chi_f$ in the case of massive quarks can also be solved numerically. 
In Fig.~\ref{chisolutionzphys}, we display
the solutions of $\chi_f$ with the physical quark masses, $m_{l}=3.45 \text{ MeV}$, $m_{s}=93.4 \text{ MeV}$, and $m_{c}=1.27 \text{ GeV}$~\cite{ParticleDataGroup:2022pth}, 
where the temperature is taken as $T=0.150$~GeV.
This figure clearly shows that the solutions $\chi_f$ take different values owing to
the violation of the $SU(4)$ quark flavor symmetry. 
This violation is reflected in the quark condensates, $\sigma_{l}=(0.3552 \text{ GeV})^{3}$, $\sigma_{s}=(0.341 \text{ GeV})^{3}$, and $\sigma_{c}=(0.3128 \text{ GeV})^{3}$.

To verify the consistency of the model, the Gell-Mann-Oakes-Renner (GOR) relation is used, $f_{\pi}^{2} m_{\pi}^{2} = 2 m_{q} \sigma$. The value of the light quark condensate at zero temperature for physical quark masses is obtained.
With the values of the light quark mass and condensate, the pion mass and the pion decay constant are evaluated (see Ref. \cite{Ahmed:2023zkk} for details on calculating the meson spectra and decay constants). 
For the physical quark mass, we have the light quark condensate $\sigma_{l}=0.07095 \text{ GeV}^{3}$,
the pion mass 
$m_{\pi}=0.17957$ GeV and the pion decay constant $f_{\pi}=0.12314$ GeV.
Actually, these values satisfy the GOR relation.

%%%%%%%%%%%%%%%%%%%%%%%%%%%%%%%%%%%%%%%%%%%%
%%%%%%%%%%%%%%%%%%%%%%%%%%%%%%%%%%%%%%%%%%%%
\section{Chiral phase transition with four-quark flavors}
\label{sectionIII}

With the above setup, we are now ready to explore the chiral phase transition including the four-quark flavors. 

 In this section, we will discuss the temperature dependence of the quark condensate and the order of the chiral phase transition.
This study is based on the soft wall holographic QCD model in the four-flavor framework with $N_f=4$.
By taking heavy quark mass limit for the strange and/or charm quarks, we can address the 
two- and three-quark flavor system. 
At the beginning of this section, we will study the order of the chiral phase transition at the physical quark mass for the three- and four-quark flavor system, respectively. To validate our approach, we will also explore the chiral restoration for the two-quark flavor system. With the results of the order of the chiral phase transition, we will describe the conventional Colombia plot for the three-quark flavor system and investigate the charm quark mass contribution to the Colombia plot. 
Towards the end of this section, we will propose the extended Columbia plot for the four-quark flavor system where the light- and strange-quark masses are degenerate
but differ from the value of the charm quark mass. Furthermore, 
we will also provide the critical exponents related to the chiral phase transition for two-, three- and four-quark flavor system.

%%%%%%%%%%%%%%%%%%%%%%%%%%%%%%%%%%%%%%%%%%%%%%%%%%%%%%%%%%%%%%%%%%%%%%%%%%%%%%%%
\subsection{
Chiral crossover with physical quark masses
in three- and four-quark flavor system
}
\label{TCandTpc}

In this subsection, we discuss the chiral phase transition at physical quark masses with three- and four-quark flavors.
To exhibit the temperature dependence on the light quark condensate (chiral condensate),
we take the following values of the physical quark masses in the case of the four-quark flavors: $m_{l}=3.45 \text{ MeV}$, $m_{s}=93.4 \text{ MeV}$, and $m_{c}=1.27 \text{ GeV}$~\cite{ParticleDataGroup:2022pth}.
On the other hand, to get from the soft-wall model with $N_f=4$ to the three-quark flavor system, 
the value of the charm quark mass is changed to heavy one which is taken as $m_{c}=3 \text{ GeV}$ in our study, while the light- and strange-quark masses are fixed as the physical values. 

\begin{figure}
\begin{subfigure}{0.49\textwidth}
  \centering
  \includegraphics[width=1\linewidth]{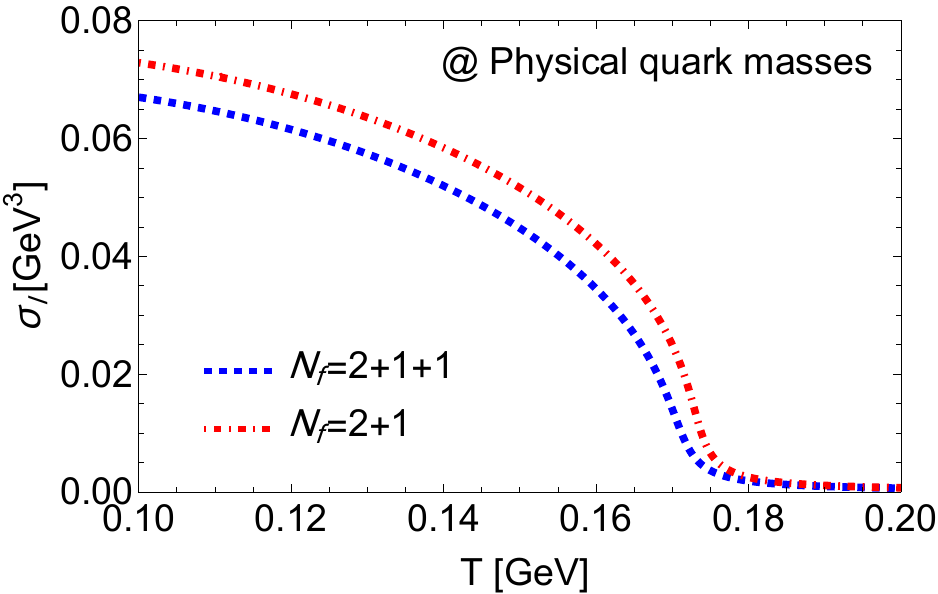} 
   \caption{ }
\label{dsigma43a}
\end{subfigure} 
\begin{subfigure}{0.473\textwidth}
  \centering
  \includegraphics[width=1\linewidth]{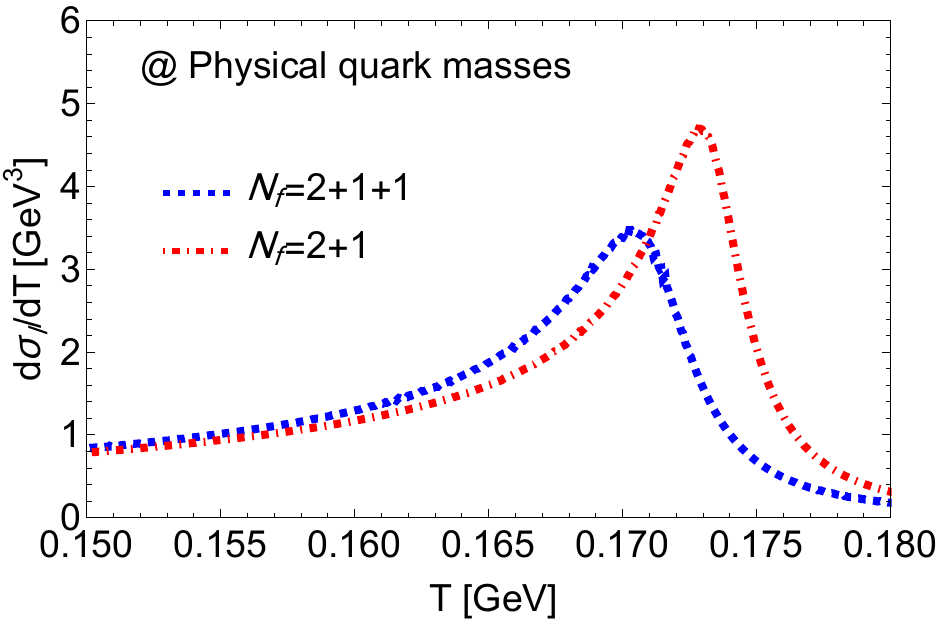}
 \caption{ }
\label{dsigma43b}
\end{subfigure}
\caption{Panel (a): The light quark condensate behavior for the four-quark flavors
(dashed blue line denoted with $N_{f}=2+1+1$) and the three-quark flavors
(dot-dashed red line with $N_{f}=2+1$) with the physical quark masses. Panel (b): The variation of $\sigma_{l}$ with respect to temperature $T$ for similar case as panel (a). The maximum points correspond to the pseudo-critical temperatures ($T_{\rm pc}$).  
The pseudocritical temperatures are $T_{c}=0.1703$~GeV and $0.1729$~GeV for the four- and three-quark flavors, respectively. }
\label{dsigma43}
\end{figure}

In Fig.~\ref{dsigma43}, we plot the light quark condensate $\sigma_{l}$ as a function of temperature in the case of the three- and four-quark flavor system with the physical quark masses.
The panel~(a) shows that the chiral condensate smoothly decreases with the increase of the temperature, but does not take exact zero even at high temperatures. This implies that 
the holographic QCD model in Eq.~(\ref{action}) undergoes the chiral crossover and the chiral symmetry is approximately restored at finite temperatures. 

The chiral crossover is characterized by the pseudocritical temperature which is evaluated by the reflection point of the quark condensate, $d^2 \sigma_l/dT^2 \bigl|_{T=T_{\rm pc}}=0$.
The panel~(b) of Fig.~\ref{dsigma43} shows the variation of $\sigma_{l}$ with temperature, which has the peak structure. 
The temperature at the peak point actually corresponds to the pseudocritical temperature  $T_{\rm pc}$.
For the three-quark flavor system with physical masses of light and strange quarks, 
the pseudocritical temperature is evaluated as  $T_{\rm pc}\bigl|_{\rm hQCD} = 0.1729~{\rm GeV}$. 
This value is smaller than the conventional effective model results like the NJL evaluation: $T_{\rm pc}\bigl|_{\rm NJL}\simeq 0.189\,{\rm GeV}$~\cite{Cui:2021bqf}.
On the other hand, the estimate of the holographic QCD is larger than the lattice observation of the three-flavor QCD with the physical quark masses, 
$T_{\rm pc}\bigl|_{\rm lat.}\sim 0.1565\,{\rm GeV}$~\cite{HotQCD:2018pds}. 
In comparison with the lattice QCD observation, 
there exists a $10\%$ deviation. 
However, the holographic QCD model is defined at the large-$N_c$, so that the deviation by about $30\%$ would be acceptable within the large-$N_c$ approximation\footnote{ In the AdS/QCD correspondence from one side there is a classical supergravity theory, and from the other side there is a gauge theory at the large $N_c$ limit (only the leading term in the large $N_c$ expansion contribute to the correspondence). In order to apply the  AdS/QCD correspondence to the real world QCD physics, we need to extrapolate the number of color to $N_c=3$, and this extrapolation requires to add the corrections to the gravity side (quantum correction), which can be be obtained as loop effects when one replaces the classical action with an effective action \cite{Gubser:1998bc}. The leading order correction to the supergravity action is in order of $\mathcal{O}(1/N_{c})$. Then, one can naively expect that the theoretical uncertainty is around $30\%$ \cite{Forkel:2010gu,Cappiello:2015baa,Richardson:2022hyj}. Unfortunately, the evaluation of the next to leading order of the $1/N_c$ expansions is not established in holographic QCD.}. 
Thus, our result qualitatively aligns well with the lattice observation.

We also evaluate the pseudocritical temperature for the four-quark flavor system with the physical quark masses, which is $T_{\rm pc}\bigl|_{\rm hQCD} = 0.1703\,{\rm GeV}$. 
Although the physical charm quark mass reduces the pseudocritical temperature, the decrement is relatively small.

%%%%%%%%%%%%%%%%%%%%%%%%%%%%%%%%%%%%%%%%%%%%
\subsection{
Two-flavor chiral phase transition from the four-flavor framework
}

When the quark masses deviate from their physical values, the chiral crossover would change to the first or second order phase transition, as was observed in the previous holographic studies \cite{Colangelo:2011sr,Cui:2013zha,Bartz:2016ufc,Chelabi:2015cwn,Chelabi:2015gpc,Fang:2015ytf,Fang:2016nfj,Li:2016smq,Bartz:2017jku}. 
Using the soft-wall model with $N_f=4$, we explore the influence of quark masses on the chiral phase transition. 
To facilitate our discussion on the chiral restoration, we focus on the two-quark flavor system in this subsection, 
which is effectively realized by taking the heavy quark mass limit for the strange and charm quarks.

\begin{figure}
\begin{subfigure}{0.49\textwidth}
  \centering
  \includegraphics[width=1\linewidth]{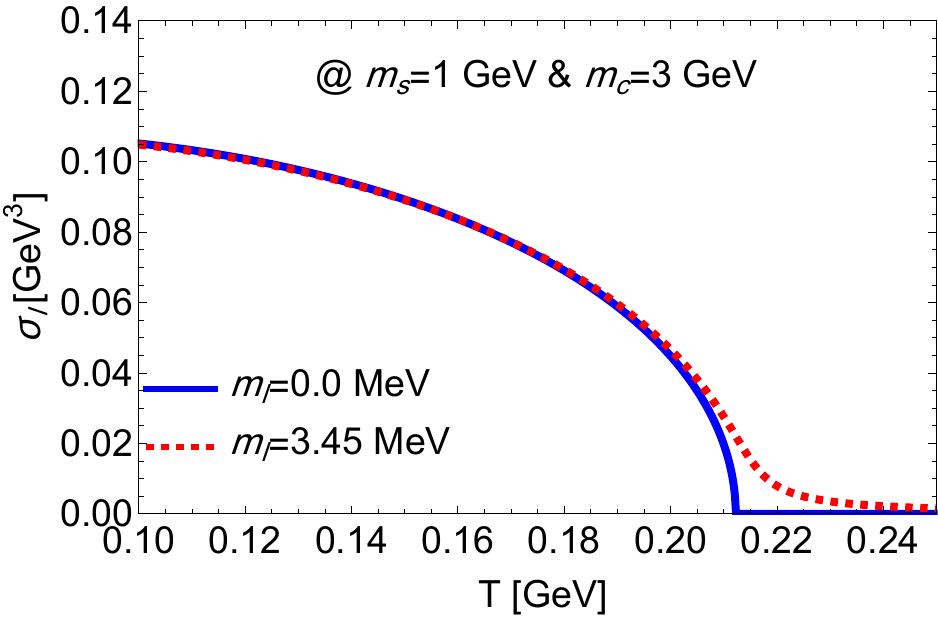} 
   \caption{ }
\label{dsigma2a}
\end{subfigure} 
\begin{subfigure}{0.49\textwidth}
  \centering
  \includegraphics[width=1\linewidth]{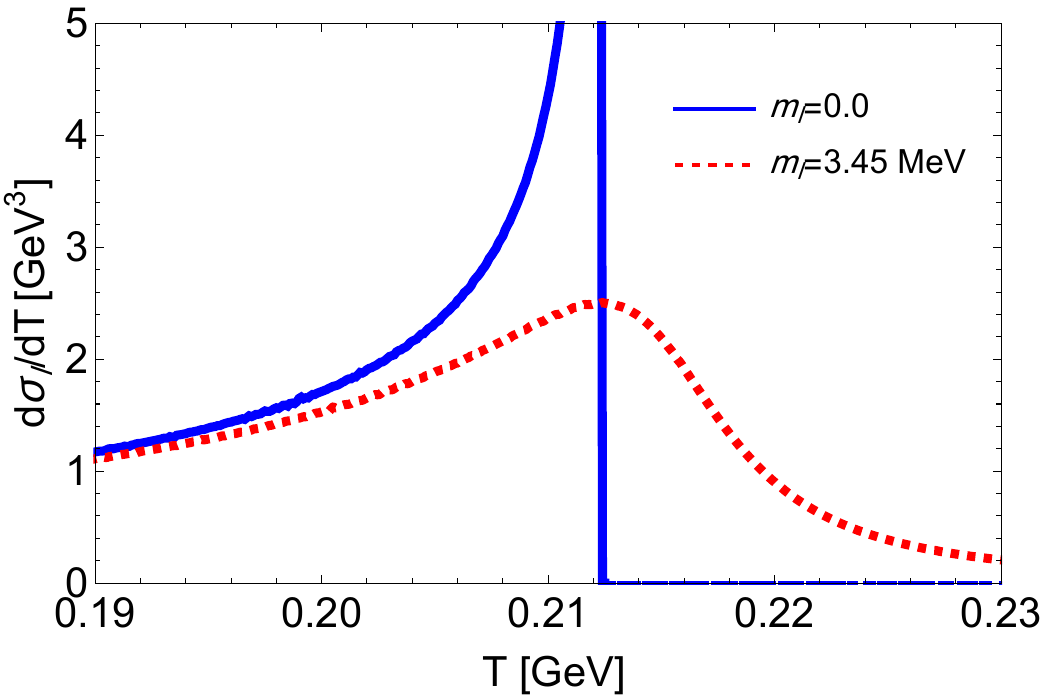}
 \caption{ }
\label{dsigma2b}
\end{subfigure}
\caption{Panel (a): Temperature dependence of the light quark condensate with 
the large strange quark mass $m_{s}=1$~GeV and charm quark mass $m_{c}=3$~GeV. This case corresponds to the two-quark flavor system for the massless (solid blue line) and physical quark mass (dashed red line) of the light quark. Panel (b): The variation of $\sigma_{l}$ with temperature $T$ with similar color coding as panel~(a). }
\label{dsigma2}
\end{figure}

In Fig.~\ref{dsigma2}, we show the influence of the light quark mass on the chiral phase transition for the two-quark flavor system, where the strange- and charm-quark masses are taken as $m_s=1$~GeV and $m_c=3$~GeV to be effectively integrated out from the four-flavor framework.  
Like in the case of the three- and four-quark flavor system, we observe the chiral crossover with the physical value of the light quark mass, as depicted in the panel~(a) of Fig.~\ref{dsigma2}. 
The pseudocritical temperature is determined to be $T_{\rm pc}=0.2124$~GeV from the panel~(b) of Fig.~\ref{dsigma2}.
When comparing the cases of the three- and four-quark flavor system, the two-flavor pseudocritical temperature is higher.
This shows that the inclusion of the strange quark significantly reduces the pseudocritical temperature in both the three- and four-quark flavor systems.

When taking the chiral limit ($m_l=0$), the chiral crossover is changed to the chiral second-order phase transition.  
The critical temperature of the second order phase transition reads $T_{\rm c}=0.2123$~GeV, determined from the point of the divergence in the panel~(b) of Fig.~\ref{dsigma2}. After reaching the critical temperature, the chiral condensate becomes zero, so that the chiral symmetry is completely restored at high temperatures, $T>T_{\rm c}$. 
Moreover, this second order is anticipated to belong to the $O(4)$ universality class~\cite{Pisarski:1983ms}.

Indeed, the second order phase transition has also been observed in the previous study for the massless two-quark flavor system based on the soft-wall model with $N_f=2$ and $N_f=3$ (where strange quark is taken to be heavy)~\cite{Chelabi:2015cwn,Chelabi:2015gpc}. Here, our numerical analysis certainly shows that the soft-wall model with $N_f=4$ can effectively replicate the two-flavor chiral phase transition by taking the heavy mass limit of the strange and charm quarks.

%%%%%%%%%%%%%%%%%%%%%%%%%%%%%%%%%%%%%%%%%%%%
\subsection{
Three-flavor first order phase transition and
influence of charm quark mass
}

The previous studies based on the soft-wall model with $N_f=3$ have also provided the chiral first-order phase transition in the case of the massless three-quark flavors and have shown that the first order phase transition is driven by 
the quark-flavor symmetry breaking via 
the determinant term $V_{\rm det}$ tagged with the parameter~$v_{\rm det}$~\cite{Chelabi:2015cwn,Chelabi:2015gpc}. 
To address the first order phase transition in the framework of the soft-wall model with $N_f=4$, we move onto the three-quark flavor system in this subsection. 
In addition, we also consider the influence of the finite charm quark mass on the three-flavor first order phase transition. 

\begin{figure}
\begin{subfigure}{0.48\textwidth}
  \centering
  \includegraphics[width=1\linewidth]{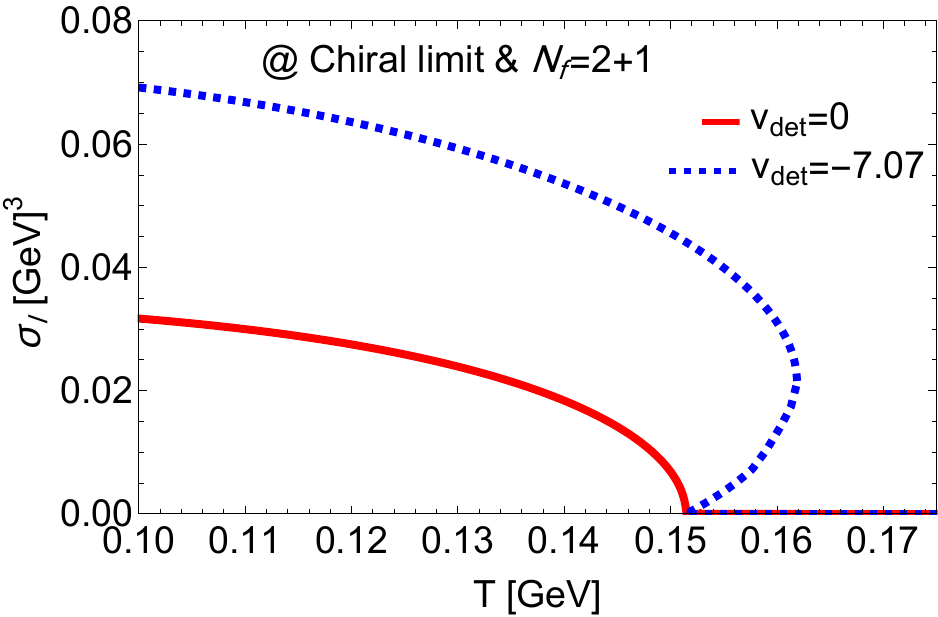} 
   \caption{ }
\label{sigmaT3yaxisa}
\end{subfigure} 
\begin{subfigure}{0.49\textwidth}
  \centering
  \includegraphics[width=1\linewidth]{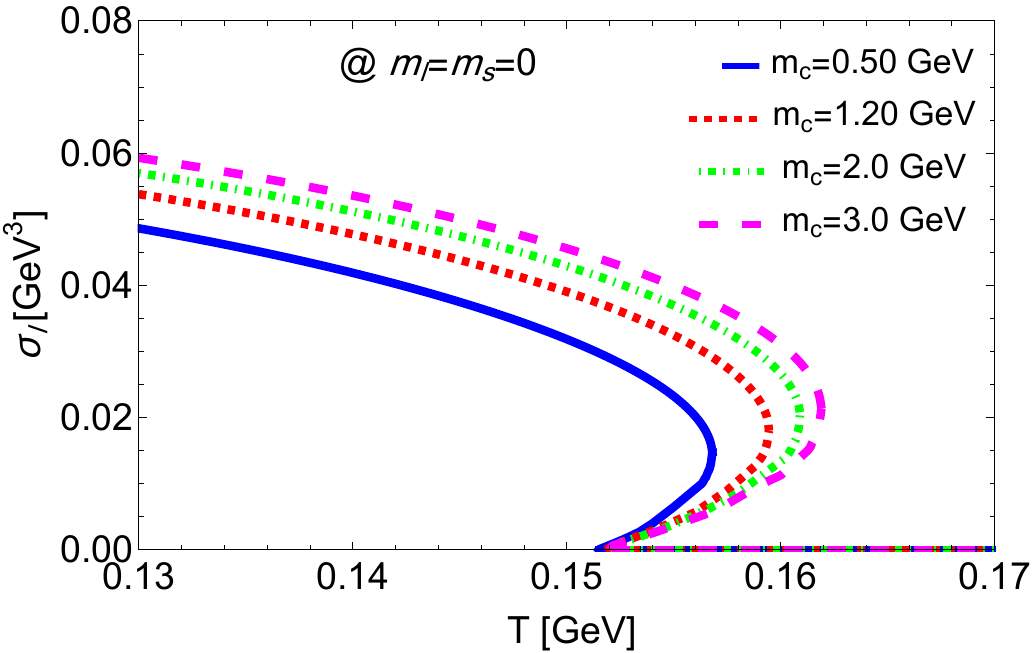} 
   \caption{ }
\end{subfigure} 
\caption{Panel (a): The light quark condensate as a function of temperature without ($v_{det}=0$) and with ($v_{det}=-7.07$) the contribution of the determinant term. Panel (b): Light quark condensate as a function of temperature for a fixed value of $m_{l}=m_{s}=0$ and changing charm quark mass as $m_{c}=0.50$ GeV (solid blue line), $m_{c}=1.20$ GeV (dashed red line), $m_{c}=2$ GeV (dash-dotted green line), and $m_{c}=3$ GeV (dashed magenta line). }
\label{sigmaT3yaxis}
\end{figure}

The panel~(a) of Fig.~\ref{sigmaT3yaxis} shows 
the contribution of the parameter~$v_{\rm det}$ on the chiral phase transition. Here, we set the massless three-quark flavor system where the charm quark mass takes the heavy value ($m_c=3$~GeV) to be integrated out.  
This figure clearly shows that the presence of the determinant term surely induces the first order phase transition even when starting from the framework of the soft-wall model with $N_f=4$. 

Next, adjusting the charm quark mass to be small, we consider the influence of the charm quark mass on the first order phase transition.
The panel~(b) shows that the critical temperature is moved to the low temperature region with the decrease of $m_c$, while the first order phase transition is intact for the finite charm quark mass. 

\begin{figure}
\begin{subfigure}{0.49\textwidth}
  \centering
  \includegraphics[width=1\linewidth]{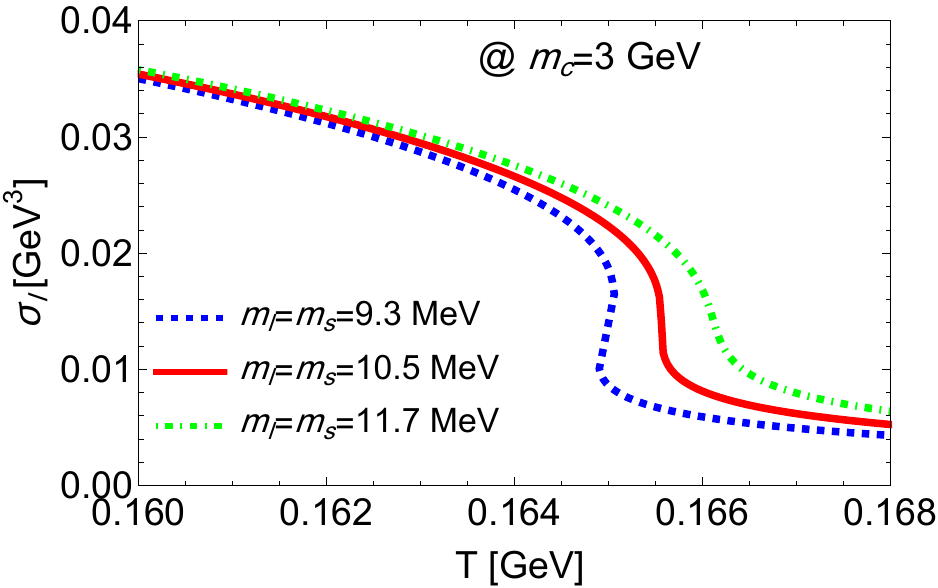}
 \caption{ }
\label{sigmaT3+1yaxisa}
\end{subfigure}%
\begin{subfigure}{0.499\textwidth}
  \centering
  \includegraphics[width=1\linewidth]{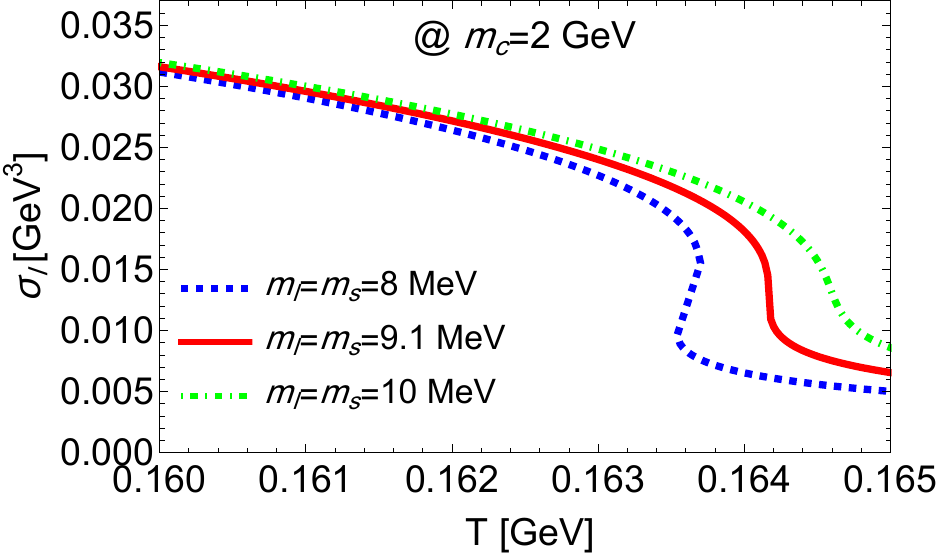}
 \caption{ }
\label{sigmaT3+1yaxisb}
\end{subfigure}%

\begin{subfigure}{0.49\textwidth}
  \centering
  \includegraphics[width=1\linewidth]{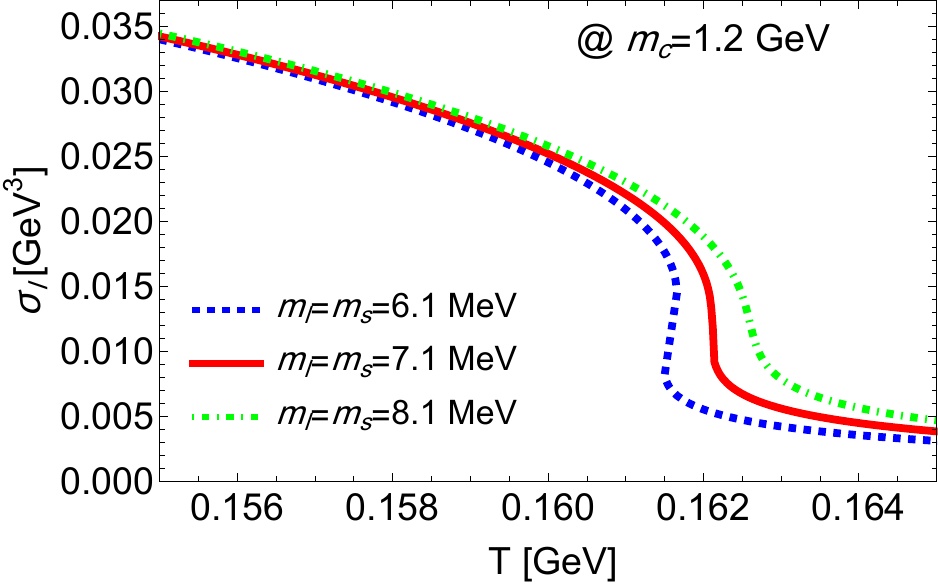}
 \caption{ }
\label{sigmaT3+1yaxisc}
\end{subfigure}%
\begin{subfigure}{0.499\textwidth}
  \centering
  \includegraphics[width=1\linewidth]{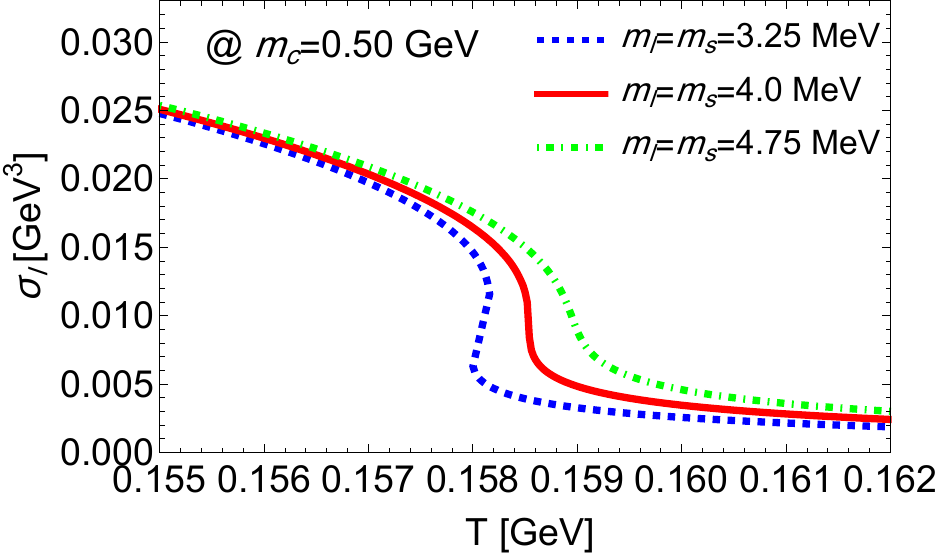}
 \caption{ }
\label{sigmaT3+1yaxisd}
\end{subfigure}%
\caption{Panel (a): The temperature dependence  of the light quark condensate for a fixed value of $m_{c}=3$ GeV and changing the degenerate light and strange quark masses to see how the order of phase transition changes by changing the mass, as $m_{l}=m_{s}=9.3$ MeV (dashed blue line), $m_{l}=m_{s}=10.5$ MeV (solid red line), and $m_{l}=m_{s}=11.7$ MeV (dash-dotted green line). The color online for panel (b), (c), and (d) is similar to panel (a) with different quark masses. }
\label{sigmaT3+1yaxis}
\end{figure}

Turning on the mass of the light quark, the chiral symmetry 
is explicitly broken and is not completely restored even at high temperatures. Then, the first order phase transition is contaminated by the current quark mass and makes the phase transition crossover at the sufficient value of the quark mass. As for the case of the $m_c=3$~GeV (see the panel (a) of Fig.~\ref{sigmaT3+1yaxis}),
the order of the phase transition is altered at $m_l=m_s\sim 10$~MeV. 
This quark mass value represents the critical quark mass to separate the region between the first-order and crossover domains.
As the charm quark mass decreases, the critical quark mass is shifted to be small, as depicted in other panels of Fig.~\ref{sigmaT3+1yaxis}:
the critical masses are evaluated as $m_l=m_s\sim 10.5$~MeV, $m_l=m_s\sim 9$~MeV, $m_l=m_s\sim 7$~MeV and $m_l=m_s\sim 4$~MeV for $m_c=3$~GeV, $m_c=2$~GeV, $m_c=1.2$~GeV and $m_c=0.5$~GeV, respectively.

\begin{figure}
  \centering
  \includegraphics[width=0.5\linewidth]{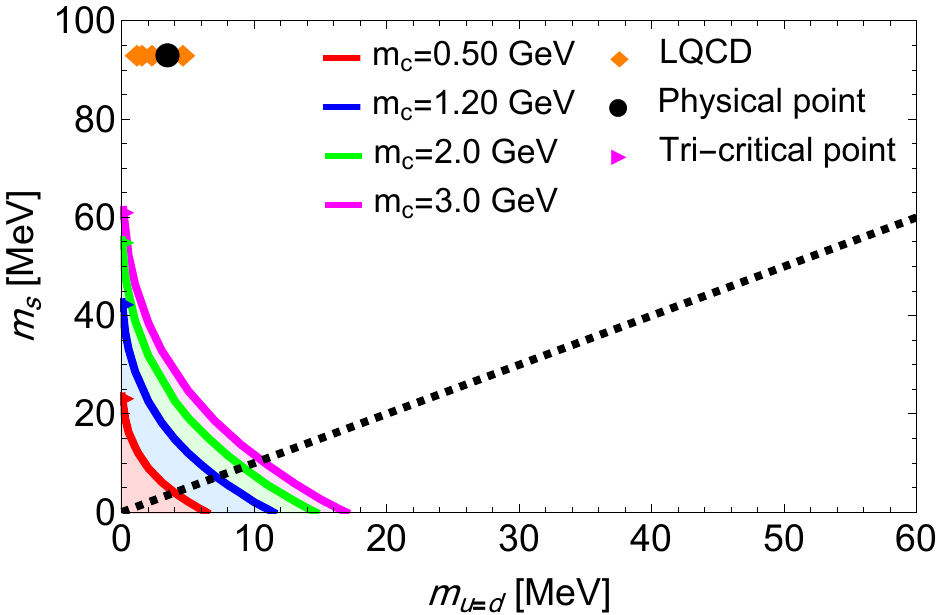}   

\caption{The phase diagram for chiral phase transition in $m_{l}-m_{s}$ plane with four different values of charm quark mass. The solid red line shows the critical line (second order line) between the first order region (the bottom left corner) and the crossover region(the upper right corner) at $m_{c}=0.50$ GeV. Similar critical line is shown for $m_{c}=1.20$ GeV (solid blue line), $m_{c}=2.0$ GeV (solid green line), and $m_{c}=3.0$ GeV (solid magenta line). 
Note that there exists the second order line in the $m_s$ axis above the critical value of the strange quark mass corresponding to the tri-critical point.
The white region represents the crossover domain and the $m_l$ axis above the critical value of the light quark mass also falls within the crossover domain. The orange points are lattice QCD results taken from Ref.~\cite{HotQCD:2019xnw,Ding:2019fzc} which are in the crossover region.
}
\label{Hcolombia2+1}
\end{figure}

To grasp the quark mass dependence on the chiral phase transition, we map the order of the chiral phase transition on a phase diagram in the quark-mass plane of $m_{l}$ and $m_{s}$, as shown in Fig.~\ref{Hcolombia2+1}. 
This figure is a sort of the Columbia plot, in which the charm quark mass contribution is included. Here, we have considered the four different values of the charm quark mass. 
The four solid lines corresponding to each charm quark mass represent the critical quark masses of $m_{l}$ and $m_{s}$. The colored domain denotes 
the region where the first order phase transition occurs, while the white regions signify the occurrence of the chiral crossover.

In the three-quark flavor system, corresponding to the case of $m_c=3$~GeV, 
the first order domain is distant from the physical point. 
Such a narrow first order domain has been also reported in 
a mean-field level analysis of the NJL model~\cite{Fukushima:2008wg} and 
the FRG method with the constant anomaly coupling~\cite{Resch:2017vjs}.
%and is not ruled out by the lattice observations.

As the charm quark mass decreases, the charm quark contribution becomes significant in the thermal system. The phase diagram in Fig.~\ref{Hcolombia2+1} shows that the first order domain becomes shrink by adjusting $m_c\searrow 0$. The first order domain is also contaminated by the charm quark contribution.
This contamination implies that the first order domain vanishes in the massless limit of the charm quark. The fate of the first order domain in the four-flavor system with the massless charm quark will be discussed in the subsequent subsection.

%%%%%%%%%%%%%%%%%%%%%%%%%%%%%%%%%%%%%%%%%%%%
\subsection{
Emergence of second order phase transition and four-flavor phase diagram
}

\begin{figure}
  \centering
  \includegraphics[width=0.5\linewidth]{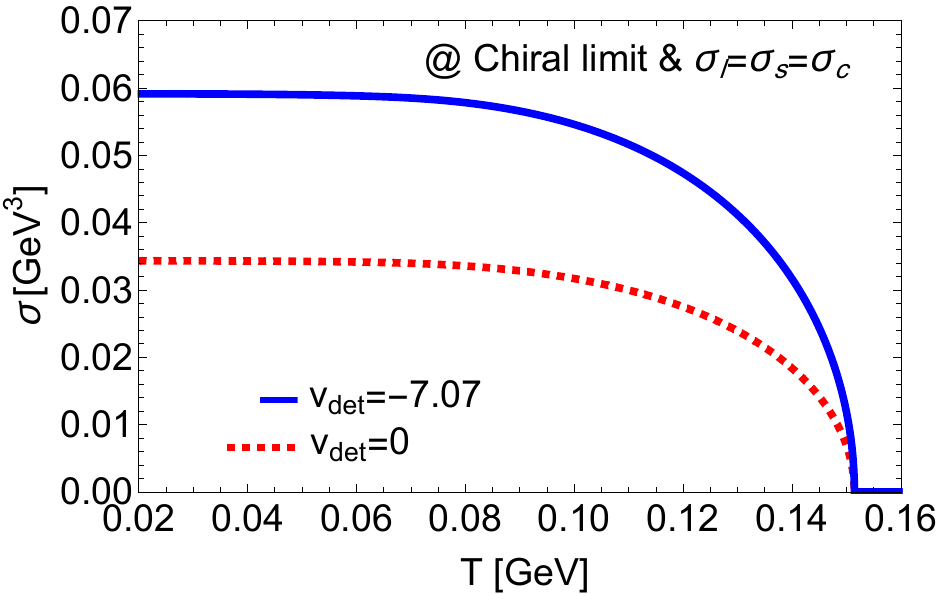} 
\caption{The quark condensate as a function of temperature for the massless quarks at  $v_{det}=-7.07$ (solid blue line) and $v_{det}=0$ (dashed red line).}
\label{sigmaTmassless}
\end{figure}

Moving onto the massless four-quark flavor system, we make a plot of the chiral phase transition in a way similar to the case of the massless three-flavor system in the panel~(a) of Fig.~\ref{sigmaT3yaxis}.
In this case, the second order phase transition emerges,
as shown in Fig.~\ref{sigmaTmassless}.
The figure indicates that
the presence of the determinant term is irrelevant to 
the second order phase transition.
This is in contrast to the massless three-flavor system depicted in the panel~(a) of Fig.~\ref{sigmaT3yaxis}, and the critical temperature is not influenced by the determinant term contribution, holding steady at $T_{c}=0.1515$~GeV.

As shown below, 
the mechanism of the second order phase transition with the massless four-quark flavors is analogous to that with the massless two-quark flavors. In the flavor-symmetric system for $N_f=4$, where $\chi_l=\chi_s=\chi_c$, the potential $V(X)$ is expressed as
\begin{eqnarray}
V(\chi_l)= a \chi_l^2+b\chi_l^4,
\end{eqnarray}
with $a = -6$ and $b= 4 v_4+3 v_{\rm det}$.
On the other hand, the potential for $N_f=2$ goes like
\begin{eqnarray}
V(\chi_l)= \bar a \chi_l^2+\bar b\chi_l^4,
\end{eqnarray}
where $\bar a$ and $\bar b$ are give by
$\bar a  =-6 + \bar v_{\rm det} $ and 
$\bar b = 2v_4$ with $\bar v_{\rm det} = \gamma/2$ being the parameter of the determinant term for $N_f=2$.
The potentials in the soft-wall model exhibit the same structure for both the two- and four-quark flavors: the potentials are constructed by only the $\chi_l^2$ and $\chi_l^4$ terms.
In these cases, the contribution of the determinant term becomes invisible in the potential structure. This is because the determinant term for $N_f=4$ ($N_f=2$) is incorporated into the $\chi_l^4$ term (the $\chi_l^2$ term).
Owing to the similarity to the potential form in the massless two-quark flavor system,
the second order phase transition also manifests in the massless four-flavor system. 
Moreover, this similarity would indicate that the universality class of the second order in the massless four-quark flavors may align with that observed in the massless two-quark flavors.

Furthermore, the steady behavior of the critical temperature for $N_f=4$ can also be understood as analogous to the observations in the two-flavor system. The previous study has shown that the critical temperature with the massless two-quark flavors remains unaffected by the choice of $v_4$ and is instead determined by only the parameter set in the dilaton profile \cite{Chelabi:2015gpc}. 
Given this fact, the critical temperature with the massless four-quark flavors 
is not shifted by the contribution of the determinant term because $v_{\rm det}$ is absorbed into the parameter of the $\chi_l^4$ term.

\begin{figure}
  \centering
  \includegraphics[width=0.495\linewidth]{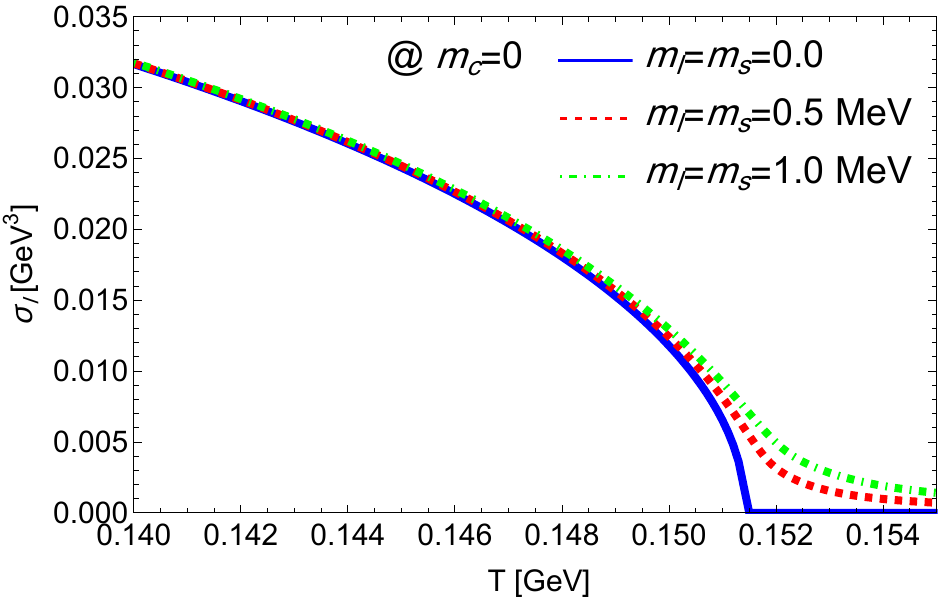}   
\caption{
Light quark condensate as a function of temperature for massless charm quark and different values of the light and strange quark masses as $m_{l}=m_{s}=0$ (solid blue line), $m_{l}=m_{s}=0.5$ MeV (dashed red line), and $m_{l}=m_{s}=1$ MeV (dash-dotted green line). 
}
\label{sigmaTmasslesscharm}
\end{figure}

When the finite mass of the light quark is introduced in the four-quark flavor system, 
the second order phase transition is contaminated, similar to the case of the first order phase transition in the three-quark flavor system shown in Fig.~\ref{sigmaT3+1yaxis}.
Figure~\ref{sigmaTmasslesscharm} clearly illustrates that the light quark mass makes the second order phase transition crossover. 
In contrast to the three-flavor first order phase transition,
this system promptly undergoes the chiral crossover as soon as the quark mass becomes massive, leading to the absence of a critical mass of the light quark in the case of  $m_c=0$.

\begin{figure}
  \centering
  \includegraphics[width=0.55\linewidth]{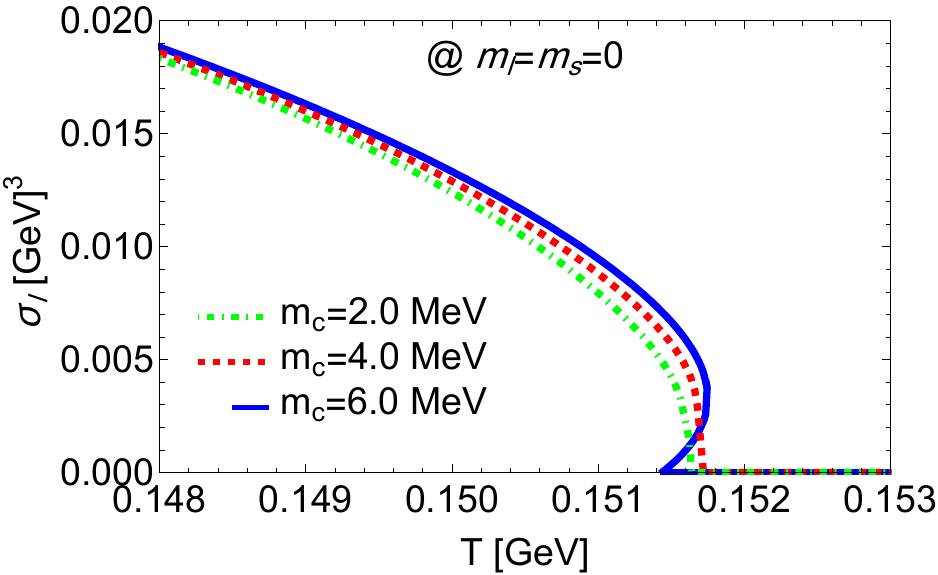} 
\caption{The light quark condensate as a function of temperature for a massless light and strange quarks $m_{l}=m_{s}=0$ and changing charm quark mass to see how the order of phase transition changes by changing the charm quark mass as $m_{c}=2$ MeV (dash-dotted green line), $m_{c}=4$ MeV (dashed red line), and $m_{c}=6$ MeV (Solid line). The critical charm quark mass that changes the order of the chiral phase transition from second order to first order is between $4-6$ MeV. }
\label{sigmaT3yaxisc}
\end{figure}

Next, we move onto the case of the massive charm quark from the massless four-quark flavor system.
Figure~\ref{sigmaT3yaxisc} illustrates that as the charm quark mass increases, the second order phase transition shifts to the first order phase transition around $m_c\sim 5$~MeV.
This value represents the critical charm quark mass that distinguishes between the crossover and the first-order phase transition.

\begin{figure}
  \centering
  \includegraphics[width=0.5\linewidth]{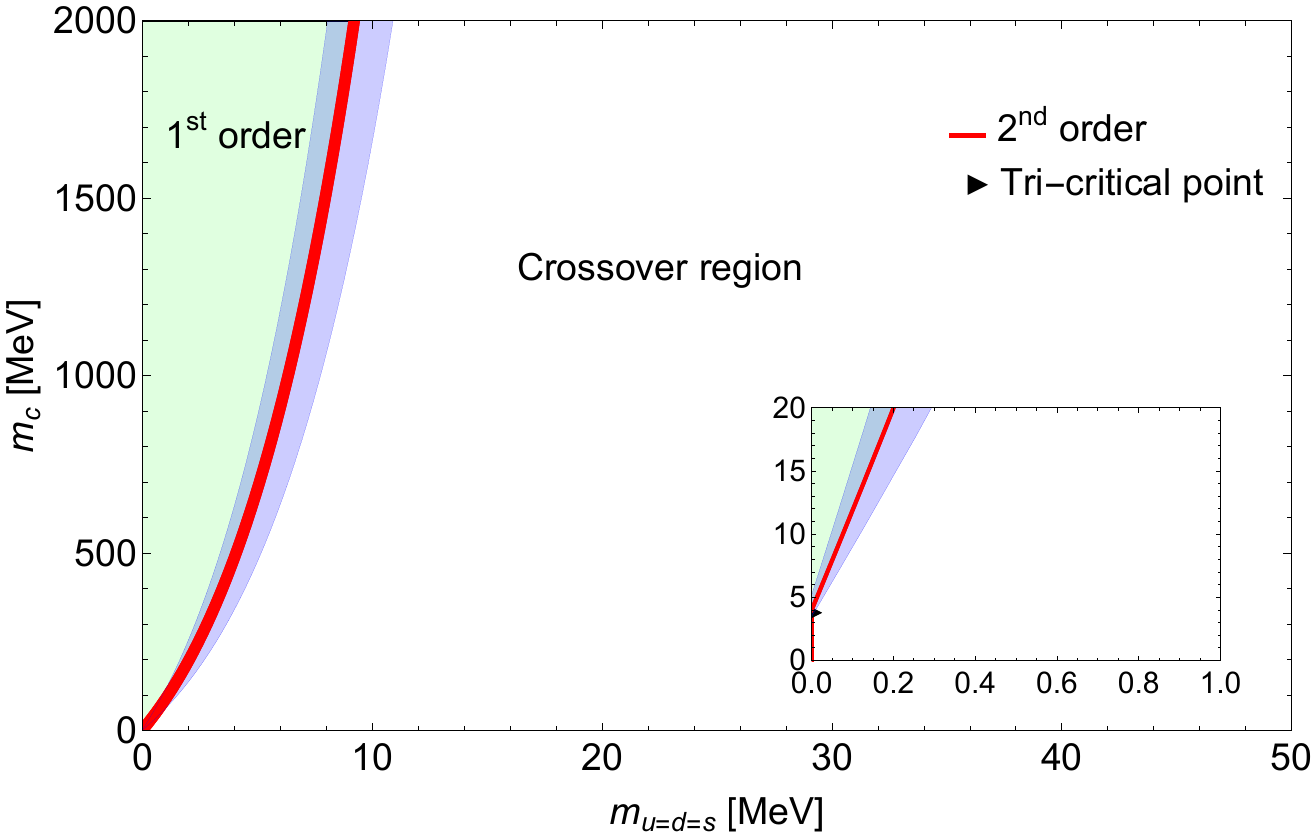}
\caption{The phase diagram for chiral phase transition in $m_{l=s}$-$m_{c}$ plane. The solid red line shows the critical line (second order line) between the first order region(light-green shaded area) and the crossover region(the white region). The tri-critical point is located at $m_{c}\sim 4$ MeV for which the order of chiral phase transition changes from second order to first order phase transition. Below the tri-critical point, there is only second order phase transition on the y-axis which is at massless $m_{l=s}=0$ and the order changes to crossover for any deviation from the massless limit. The blue band corresponds to the naive $30\%$ corrections from the large $N_c$ estimation. }
\label{colombia3+1}
\end{figure}

By combining the results of Figs.~\ref{sigmaTmassless}, \ref{sigmaTmasslesscharm} and  \ref{sigmaT3yaxisc}, 
we depict the phase diagram at the four-quark flavor system in Fig.~\ref{colombia3+1} where
$m_l$ and $m_s$ are degenerate but differ from $m_c$.
This phase diagram is on the $m_{l=s}$-$m_c$ plane, which is mainly drawn by the first order domain and the crossover domain. 
The coexistence of the two domains in the phase diagram is formed by the competition between the contribution of the determinant term associated with the flavor symmetry breaking and the current quark mass source related to the explicit chiral symmetry breaking, similar to the phase diagram on the $m_l$-$m_s$ plane shown in Fig.~\ref{Hcolombia2+1}.

Furthermore, there exists the tri-critical point on the $m_{l=s}$-$m_c$ plane ($m_l$ degenerates with $m_s$), at which the three domains of crossover, first- and second-order phase transitions merge. 
Of interest is that the tri-critical point corresponds to 
the critical charm quark mass estimated in Fig.~\ref{sigmaT3yaxisc} 
and stems from the massless four-quark flavor limit where the second order phase transition observed in Fig.~\ref{sigmaTmassless} emerges.

Our proposed extended-Columbia plot would be also characterized by the critical exponents, which is evaluated the subsequent subsection.

%%%%%%%%%%%%%%%%%%%%%%%%%%%%%%%%%%%%%%%%%%%%%%%%%%%%%%%%%%%%%%%%%%%%%%%%%%%%%%%%%%%%%
\subsection{Critical temperatures with massless quarks}

\begin{figure}
  \centering
  \includegraphics[width=0.5\linewidth]{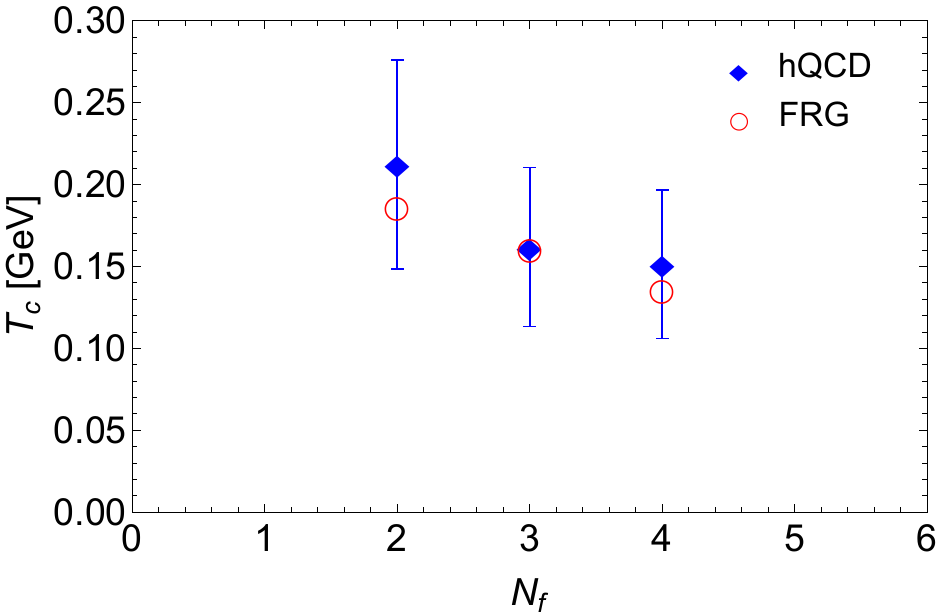}   
\caption{ The critical temperature as a function of number of flavors for massless quarks. The blue diamond is the results from our calculations, the error bar represents the $30\%$ systematic uncertainty coming from the extrapolation of the number of color to $N_c=3$, and
the red open circles are the results of the FRG model \cite{Braun:2006jd, Braun:2009ns}. } 

\label{TNf_v1}
\end{figure}

With the above results of the critical temperatures in the massless systems, we make a comparison between the holographic estimates and other model results based on the functional renormalization group (FRG) method
in Fig.~\ref{TNf_v1}. 
The FRG method~\cite{Braun:2006jd, Braun:2009ns, Braun:2023qak} provides the second order phase transition for $N_f=2,3,4$. 
However, our soft-wall model in the three-quark flavor system provides the first order phase transition while the second order phase transition arises in the two- and four-quark flavor system.
Despite the discrepancy in the phase transition order between the soft-wall model and other model results, our estimated critical temperatures can be considered to be comparable to those of other models when taking account of the large-$N_c$ approximation.

%%%%%%%%%%%%%%%%%%%%%%%%%%%%%%%%%%%%%%%%%%%%%%%%%%%%%%%%%%%%%%%%%%%%%%%%%%%%%%%%%%%%%

\subsection{Critical exponents}

This subsection delves into the criticality of chiral phase transition at finite temperature across various regions of the Columbia plot in
Figs.~\ref{Hcolombia2+1} and \ref{colombia3+1}. The scaling behavior of the quark condensate is characterized by critical exponents $\beta$ and $\delta$ near the critical point. To this end, we introduce the reduced temperature $t$, which is defined as $t=\frac{T-T_{c}}{T_{c}}$, where $T_{c}$ represents the critical temperature of the second order chiral phase transition.

The light quark chiral condensate near the critical temperature
can be defined as \cite{Rajagopal:1992qz}

\begin{equation}
    \sigma_{l} (t \to 0) \propto |t|^{\beta},  
    \label{beta}
\end{equation}

Note that the order parameter is nonzero even at the critical temperature when the value of the light quark mass takes finite values. 
This light quark mass dependence allows us to write the other critical-scaling relation
as the following,

\begin{eqnarray}
 \sigma_{l} (t,m_l ) = m_l^{1/\delta}
 f_{\sigma_l}\left(t/m_l^{1/(\delta\beta)}\right),
 \mbox{ for $m_l\sim m_{l}^{\rm cri}$},
\end{eqnarray}
where $f_{\sigma_l}$ is a function of the scaling variable $t/m_l^{1/(\delta\beta)}$
and $m_{l}^{\rm cri}$ denotes the critical mass of the light quarks at the second order of the chiral phase transition.
When the light quark mass is close to its critical value, $f_{\sigma_l}$ becomes independent of the light quark mass.
This behavior is known as the universal scaling function of the light quark mass.
Hence, 
taking $m_l \to m_{l}^{\rm cri}$ at the critical temperature, one can find the following scaling of the light quark mass,

\begin{equation}
    \sigma_{l} (t=0,m_l \to m_{l}^{\rm cri}) \propto m_l^{1/\delta}.  
    \label{delta}
\end{equation}
Note that the strange and/or charm quark mass are fixed constant values when we evaluate the critical exponent $\delta$.  

Let's start by examining the scenario for the two-quark flavor system where the masses of the strange and charm quarks are large enough to be integrated out from the action. According to Fig. \ref{dsigma2}, the critical temperature of the two flavor case at the chiral limit is $T_c=0.2122$ GeV. The value of $\beta$ and $\delta$ extracted from fitting Eq.~\eqref{beta}, and Eq.~\eqref{delta}, respectively
are $\beta = 0.50$ and  $\delta=3.0$
and we make a comparison with other models in Table~\ref{tab:exponenetNf2}. Our estimations are in line with the mean field approximation of $\beta=1/2$ and $\delta=3$ \cite{Rajagopal:1992qz}.

\begin{table}
\center
\begin{tabular}{c  c  c c c}
\hline
\hline

    Model  &    &   $\beta$ &   &   $\delta$   \\ \hline
    Mean field ($N_f=2$) \cite{Rajagopal:1992qz}   &  &   $1/2$ &   &  $3$   \\ 
    O(4) scaling theory \cite{Rajagopal:1992qz} &  &   $0.38$ &      & $4.8$  \\ 
    EMD hQCD: $N_f=2$ \cite{Zhao:2023gur}   &  &   $0.32$ &   &  $4.9$   \\ 
    hQCD: (
    $m^{\rm cri}_l=0$;
    $m_s=1$ GeV $\&$ $m_c=3$ GeV)       &  &   $0.50$ &      & $3.0$  \\ 
     \hline
     \hline
\end{tabular}
\caption{\footnotesize { The comparison of the values of the critical exponents $\beta$ and $\delta$ for the two-quark flavor system with the mean field approximation, O(4) scaling theory, Z(2) universality class, and EMD holographic QCD. }}   
\label{tab:exponenetNf2}
\end{table}

Next, we move onto the three-quark flavor system with the large charm quark mass ($m_c=3$ GeV). 
As shown in the Columbia plot in Fig.~\ref{Hcolombia2+1},
the second order line divides the chiral phase transition into two regions: the first order region and the crossover region.   
Furthermore, there exists the tri-critical point. Here, we consider three cases: case (i)  across the first order region to the crossover region for the degenerate light and strange quark mass (at the critical masses $m_l^{\rm cri}=m_s^{\rm cri}=10.5$ MeV), case (ii) at the tri-critical point with $m_{l}^{\rm tri}=0$ and $m_{s}^{\rm tri}=61.5$ MeV,
case (iii) from the second order line to the crossover region at $m_l^{\rm cri}=0$ and $m_s^{\rm cri}=100$ MeV. The results of these cases are present in Table \ref{tab:exponenetNf3}. 
Intriguingly, the critical exponents at the tri-critical point $\beta=0.25$ and $\delta=4.9$ match well with the values in the mean field approximation~\cite{Rajagopal:1992qz}. However, altering the critical values of the strange quark mass as well as the light quark mass leads to changes in the critical exponents, deviating them from the mean field values.

\begin{table}
\center
\begin{tabular}{c  c  c c c}
\hline
\hline

    Model  &    &   $\beta$ &   &   $\delta$   \\ \hline
    Mean field ($N_f=2$) \cite{Rajagopal:1992qz}   &  &   $1/2$ &   &  $3$   \\ 
    Mean field ($N_f=2+1$ @ tricritical points) \cite{Rajagopal:1992qz}   &  &   $1/4$ &   &  $5$   \\ 
    O(4) scaling theory \cite{Rajagopal:1992qz} &  &   $0.38$ &      & $4.8$  \\ 
    Z(2) universality class \cite{Campostrini:2002cf} &  &   $0.32$ &      & $4.8$  \\ 
    hQCD: ($m_{l}^{\rm cri}=m_s^{\rm cri}=10.5$ MeV;  $m_c=3$ GeV) &  &   $0.48$ &      & $2.2$  \\     
    hQCD:  ($m_{l}^{\rm tri}=0$, $m_{s}^{\rm tri}=61.5$ MeV; $m_c=3$ GeV) &  &   $0.25$ &      & $4.9$  \\
    hQCD: ($m_l^{\rm cri}=0$, $m_s^{\rm cri}=100$ MeV; $m_c=3$ GeV) &  &   $0.47$ &      & $3.2$  \\ 
     \hline
     \hline
\end{tabular}
\caption{\footnotesize {
The same as in Table.~\ref{tab:exponenetNf2} but for the three-quark flavor system.
}
}   

\label{tab:exponenetNf3}
\end{table}

Finally, we evaluate the critical exponents for the four-quark flavor system in Fig.~\ref{colombia3+1} where $m_l$ and $m_s$ are degenerate but differ from $m_c$. Similar to the three-quark flavor system,
we consider the three cases:
case (i) across the first order region to the  crossover regions at the critical masses $m_{l=s}^{\rm cri}=4$ MeV and $m_{c}^{\rm cri}=500$ MeV, case (ii) at the tri-critical point with $m_{l=s}^{\rm tri}=0$ and $m_{c}^{\rm tri}=4$ MeV,
case (iii) from the second order line to the crossover region at $m_{l=s}^{\rm cri}=0$ and $m_c^{\rm cri}=0$.
The values of the critical exponents are given in Table~\ref{tab:exponenet}. 
Our evaluation of the critical exponents at the tri-critical point would characterize the extended Columbia plot.

\begin{table}
\center
\begin{tabular}{c  c  c c c}
\hline
\hline

    Model  &    &   $\beta$ &   &   $\delta$   \\ \hline
    Mean field ($N_f=2$) \cite{Rajagopal:1992qz}   &  &   $1/2$ &   &  $3$   \\ 
    Mean field ($N_f=2+1$ @ tricritical point) \cite{Rajagopal:1992qz}   &  &   $1/4$ &   &  $5$   \\ 
    O(4) scaling theory \cite{Rajagopal:1992qz} &  &   $0.38$ &      & $4.8$  \\ 
    Z(2) universality class \cite{Campostrini:2002cf} &  &   $0.32$ &      & $4.8$  \\ 
    hQCD:  ($m_{l=s}^{\rm cri}=4$ MeV, $m_{c}^{\rm cri}=500$ MeV) &  &   $0.43$ &      & $2.6$  \\ 
    hQCD:  ($m_{l=s}^{\rm tri}=0$, $m_{c}^{\rm tri}=4$ MeV) &  &   $0.41$ &      & $3.6$  \\ 
    hQCD:  ($m_{l=s}^{\rm cri}=0$,  $m_c^{\rm cri}=0$) &  &   $0.46$ &      & $3.1$  \\
     \hline
     \hline
\end{tabular}
\caption{\footnotesize {
The same as in Table.~\ref{tab:exponenetNf2} but for the four-quark flavor system.
}}   
\label{tab:exponenet}
\end{table}

\begin{figure}
  \centering
  \includegraphics[width=0.49\linewidth]{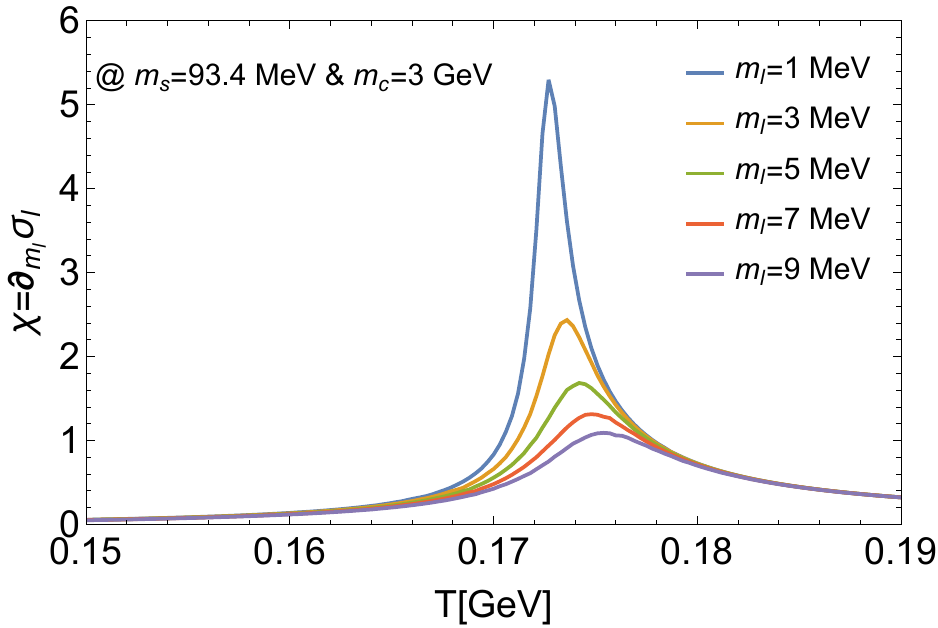}
  \includegraphics[width=0.49\linewidth]{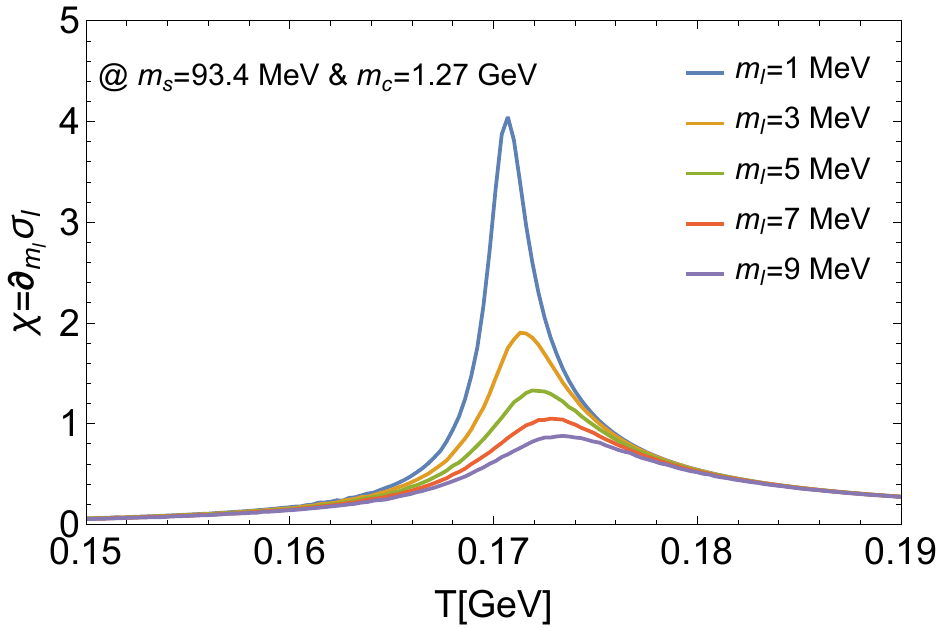}
  
   \includegraphics[width=0.49\linewidth]{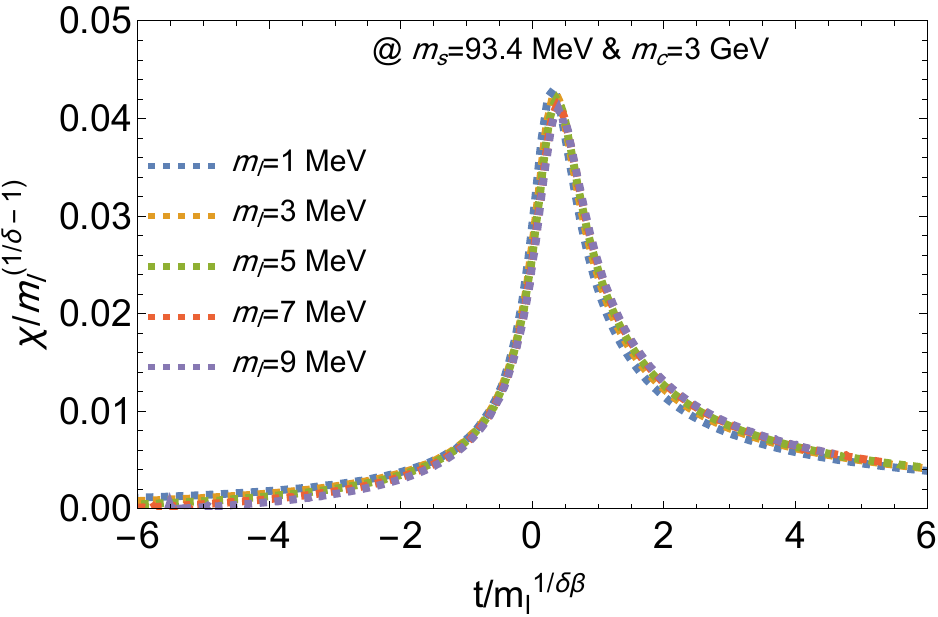}
  \includegraphics[width=0.49\linewidth]{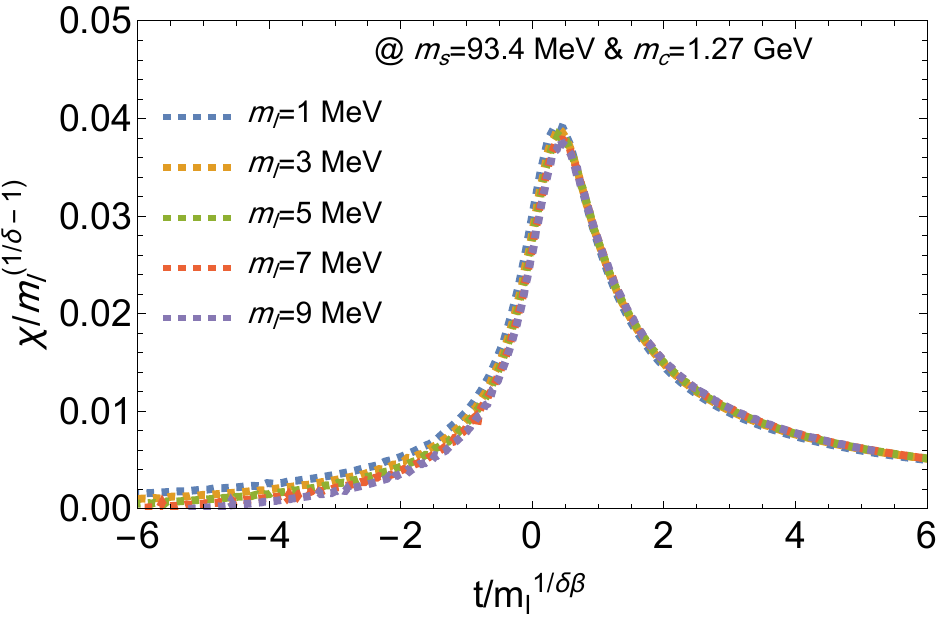}
\caption{Upper panel: Quark mass dependence of
the temperature dependence of
the chiral susceptibility with different quark masses
for three (left) and four (right) quark flavor system. Lower panel: Rescaled chiral susceptibility as the function of the scaling variable $t/m_{l}^{1/\delta \beta}$ with different quark masses
for three (left) and four (right) quark flavor system. The value of the critical exponents in the case of  $m_l^{\rm cri}=0,\;m_s^{\rm cri}=93.4\;{\rm MeV},\;
m_c=3\;{\rm GeV}$
are $\beta= 0.47$ and $\delta =3.3$. For the case of four quark flavor, $m_l^{\rm cri}=0,\;m_s^{ \rm cri}=93.4\; { \rm MeV},\;
m_c=1.27\; { \rm GeV}$, the critical exponents 
are evaluated as
$\beta=0.50$ and $\delta=3.0$.  
}
\label{susc}
\end{figure}

It is worth mention that, one can also obtain the critical exponents from the chiral susceptibility, $\chi = \partial \sigma_l/\partial m_l$.
The temperature dependence of the chiral susceptibility is plotted in Fig. \ref{susc} (upper panel). This figure shows that the chiral susceptibility has the peak structure where the temperature corresponds to the pseudocritical temperature estimated from the quark condensate. 

Moreover, taking the derivative of Eq.~\eqref{delta} with respect to the light quark mass, we can write the chiral susceptibility by using the critical exponents and scaling function:

\begin{eqnarray}
\chi(t, m_l) = 
m_l^{1/\delta -1}
 f_{\chi}\left(t/m_l^{1/(\delta\beta)}\right),
 \mbox{ for $m_l\sim m_{l}^{\rm cri}$}
\end{eqnarray}
where $f_{\chi}$ is the universal function in terms of the chiral susceptibility, which should be connected with $f_{\sigma_l}$. 
In the lower panel of Fig. \ref{susc}, we have showed $\chi/m_l^{1/\delta -1}$ as the function of the scaling variable $t/m_{l}^{1/\delta \beta}$ with different quark masses for three ($T_c=0.1721$ GeV) and four ($T_c=0.170$ GeV) quark flavor system. 
 $\chi/m_l^{1/\delta -1}$ for different values of the light quark masses are approximately collapse to one curve, identifying it as the universal function of the light quark masses $f_{\chi}$.
Note that as the light quark mass becomes large and deviates from its critical values, the collapsed curve starts to separate, implying that $\chi/m_l^{1/\delta -1}$ no longer behaves as the universal function.  
With the universal function $f_{\chi}=\chi/m_l^{1/\delta -1}$, we can read the critical exponent $\delta$ 
through $\chi(t=0,m_l\to m_l^{\rm cri}) \propto m_l^{1/\delta -1}$.
Now we consider one case where $m_s=93.4$ MeV and $m_c=3$ GeV, and then the critical exponent $\delta$ is evaluated as $\delta=3.3$. This is the same value as one obtained from Eq.~\eqref{delta} with the similar condition.

%%%%%%%%%%%%%%%%%%%%%%%%%%%%%%%%%%%%%%%%%%%%%%%%%%%%%%%%%%%%%%%
\section{Summary and discussion}
\label{sectionIIII}
In this paper, we have explored 
the quark-flavor and -mass dependence on the thermal chiral phase transition. In order to clarify the phase structure on the quark-mass plane, we have employed the soft-wall holographic QCD model with $N_f=4$, in which the quark-flavor mixing is incorporated through the determinant term interaction. 
By using the holographic model framework, the four-quark flavor system is described 
and is applicable to finite temperature across the chiral phase transition.
Furthermore, the holographic four-quark flavor system can be reduced to 
the two- or three-quark flavor system by
taking the heavy mass limit of the charm and/or strange quark.
\\

First, we have investigated the three-quark flavor system with the physical quark masses. 
In this case,
we have observed that the soft-wall model surely undergoes the chiral crossover and its pseudocritical temperature aligns well with the lattice QCD simulations within the large-$N_c$ approximation. In addition, we have also confirmed that the four-flavor model framework is capable of reproducing the observations of the phase transition order from prior holographic studies with $N_f=2,3$: in the massless limit of the two-quark  (three-quark) flavor system, the holographic model with $N_f=4$ yields the second (first) order phase transition, consistent with the prior studies~\cite{Chelabi:2015cwn,Chelabi:2015gpc}. 
This first order phase transition is driven by the contribution of the determinant term, similar to the prior studies~\cite{Chelabi:2015cwn,Chelabi:2015gpc}.
\\

Applying our framework to the three-quark flavor system for various values of quark masses, 
we have illustrated the chiral phase diagram on the plane of the light quark mass $m_l$ and the strange quark mass $m_s$, which is a part of the Columbia plot. 
For the region where $m_l$ and $m_s$ 
are small compared to their physical values, 
the phase diagram is dominated by the first order transition owing to the presence of the determinant term.

Moreover, we have also investigated the influence of the charm quark mass on 
the chiral phase diagram on the $m_l$-$m_s$ plane, especially the first order domain. It has been found that by adjusting the charm quark mass to be small, the first order domain becomes shrink. It implies that the first order domain disappears at the massless limit of the charm quark mass.
To further explore the disappearance of the first-order domain, we have investigated the massless four-quark flavor system, resulting in the emergence of the second-order phase transition.
The second-order phase transition emerged with massless four-quark flavors can be understood by the mechanism analogous to that observed in the massless two-quark flavor system.
Indeed, in the case of the massless quark system for both $N_f=2$ and $N_f=4$, the determinant term is absorbed into other interaction terms described by even functions of the scalar field.
Consequently, its contribution becomes irrelevant to evaluating the order of the chiral phase transition.
\\

Given the phase transition order observed in the massless four-quark flavor system, we have proposed the new phase diagram with the four-quark flavors, which is described on the $m_{l=s}$-$m_c$ plane. 
Even for the four-quark flavor system, both the first-order domain and the crossover domain coexist in the phase diagram. 
We have found that the tri-critical point exists in the $m_c$ axis, specifically at $ m_c\sim 4{\rm MeV}$ for $m_{l=s} =0$. 
This tri-critical point stems from
the massless four-quark flavor limit where  
the second order phase transition occurs.
\\

Incidentally, our estimated critical temperatures are comparable to these of FRG method
in the massless systems within the systematic uncertainty~\cite{Braun:2006jd, Braun:2009ns, Braun:2023qak}. However, the phase transition order of the holographic model, especially observed in the three-quark flavor system, is in contrast to the other model observations. 
\\

Finally, we estimate the critical exponents in our model for two-, three-, and four-quark flavor system. The estimated values of the critical exponents $\beta$ and $\delta$ for two-quark flavor system and the three-quark flavor system at the tri-critical point are consistent with the mean field approximation. 
For the four-quark flavor system, our estimated critical exponents at the tri-critical point would characterize our proposed extended-Colombia plot, which warrants further study in other model approaches.\\

Below, we provide some comments on our findings and their implications.

\begin{itemize}

\item 

As the pioneer work for the phase diagram on the quark-mass plane, the linear sigma model has predicted the first-order phase transition for the massless four-quark flavor system~\cite{Pisarski:1983ms}.  
In contrast, the soft-wall holographic QCD model exhibits the second-order phase transition. 
This estimation is consistent with the recent prospect of the lattice QCD simulations~\cite{Cuteri:2021ikv}.

%%%%%%%%%%%%%%%
\item 
The conventional NJL model also provides the first order phase transition for the massless three-quark flavors, which is triggered by the instanton induced anomaly term. Actually, the instanton induced anomaly term is described by the determinant term, which is called the KMT interaction~\cite{Kobayashi:1970ji,Kobayashi:1971qz,tHooft:1976rip,tHooft:1976snw}. 
However, the determinant term in the soft-wall model would not be directly linked with the KMT term and the instanton contribution is still unclear within the holographic QCD framework at finite temperature.
Hence, it is worth to study the role of the determinant term and find out the holographic dual associated with the instanton contribution that affect the order of the chiral phase transition.

%%%%%%%%%%%%%%%
\item 

Figure~\ref{TNf_v1} implies that by taking the extrapolation of the number of the quark flavor, the critical temperature would be zero at the large number of the quark flavor. 
Actually, other effective models have shown that the critical temperature vanishes at the large number of the quark flavor and have pointed out the appearance of the conformal window \cite{Appelquist:1996dq,Gies:2005as,Ryttov:2016ner,Antipin:2018asc,Ryttov:2017lkz,Alvares:2012kr}.
Furthermore, the conformal window has been also studied in the lattice QCD simulation at finite temperatures \cite{Cuteri:2021ikv}. 
Therefore, it is interesting that the soft-wall model approach is extended to the more multi-flavor system and is addressed to the conformal window. 
At any rate, 
our new phase diagram with the four-quark flavor in Fig.~\ref{colombia3+1}
characterized by critical exponents
is a milestone in investigating the chiral phase structure in the multi-flavor system. 

\end{itemize}

\section*{Acknowledgments}
We would like to thank Kazem Bitaghsir Fadafan, and Danning Li for the useful discussions. This work is supported in part by the National Natural Science Foundation of China (NSFC) Grant Nos: 12235016,
12221005, 12147150, 12305136 and the Strategic Priority Research Program of Chinese
Academy of Sciences under Grant No. XDB34030000,
the start-up funding from University of Chinese Academy
of Sciences (UCAS), and the Fundamental Research Funds
for the Central Universities. H. A. A. acknowledges the ”Alliance of International Science Organization (ANSO)
Scholarship For Young Talents” for providing financial support for the Ph.D. study.

%%%%%%%%%%%%%%%%%%%%%%%%%%%%%%%%%%%%%%%%%%%%

\bibliography{ref}% bibliography

 \addcontentsline{toc}{section}{References}
\end{document}